\input harvmac 
\input epsf.tex
\overfullrule=0mm
\newcount\figno
\figno=0
\def\fig#1#2#3{
\par\begingroup\parindent=0pt\leftskip=1cm\rightskip=1cm\parindent=0pt
\baselineskip=11pt
\global\advance\figno by 1
\midinsert
\epsfxsize=#3
\centerline{\epsfbox{#2}}
\vskip 12pt
{\bf Fig. \the\figno:} #1\par
\endinsert\endgroup\par
}
\def\figlabel#1{\xdef#1{\the\figno}}
\def\encadremath#1{\vbox{\hrule\hbox{\vrule\kern8pt\vbox{\kern8pt
\hbox{$\displaystyle #1$}\kern8pt}
\kern8pt\vrule}\hrule}}

\def\tvi{\vrule height 12pt depth 6pt width 0pt}
\def\tv{\tvi\vrule}

\Title{T96/083}
{{\vbox {
\bigskip
\centerline{Folding of the Triangular Lattice} 
\centerline{}
\centerline{with Quenched Random Bending Rigidity}
}}}

\bigskip

\centerline{P. Di Francesco, E. Guitter and S. Mori\footnote*{Address after July 1996:
 Department of Physics, Graduate School of Science, University of Tokyo,
 Hongo 7-3-1, Bunkyo-ku, Tokyo 113, Japan}}
\centerline{ \it Service de Physique Th\'eorique, C.E.A. Saclay,}
\centerline{ \it F-91191 Gif sur Yvette Cedex, France}

\bigskip

\phantom{AAAAAA}

\phantom{AAAAAA}

\phantom{AAAAAA}

\phantom{AAAAAA}

\bigskip

We study the problem of folding of the regular triangular lattice in
the presence of a quenched random bending rigidity $\pm K$ and a magnetic field $h$
(conjugate to the local normal vectors to the triangles).
The randomness in the bending energy can be understood as arising from a prior
marking of the lattice with quenched creases on which folds are favored. 
We consider three types of quenched randomness:
(1) a ``physical'' randomness where the creases arise from some prior
random folding; 
(2) a Mattis-like randomness where creases are domain walls of some quenched
spin system;
(3) an Edwards-Anderson-like randomness where the bending energy is $\pm K$ 
at random independently on each bond. 
The corresponding $(K,h)$ phase diagrams are determined in the hexagon 
approximation of the 
cluster variation method. Depending on the type of randomness, the system shows
essentially different behaviors.

\noindent
\Date{07/96}


\nref\NPW{{\it Proceedings of the Fifth Jerusalem Winter School}, ``Statistical Mechanics 
of Membranes and Surfaces'', edited by D.R.Nelson, T.Piran and S.Weinberg 
(World Scientific, Singapore, 1989)}.
\nref\N{D. R. Nelson, cond-mat/9502114.}
\nref\P{L. Peliti, cond-mat/9501076.}
\nref\KKN{Y. Kantor, M. Kardar and D. R. Nelson, Phys. Rev. Lett. 
{\bf 57} (1986) 791; Phys. Rev. {\bf A36}(1987) 3056.}
\nref\NP{D.R. Nelson and L. Peliti, J. Phys. France {\bf 48} (1987) 1085.}
\nref\PKN{M. Paczuski, M. Kardar and D.R. Nelson, Phys. Rev. Lett. {\bf 60} (1988) 2638.}
\nref\DG{F. David and E. Guitter, Europhys. Lett. {\bf 5} (1988) 709.}
\nref\AL{J. Aronovitz and T. Lubensky, Phys. Rev. Lett. {\bf 60} (1988) 2636.}
\nref\AGL{J. Aronovitz, L. Golubovic, and T. Lubensky, J. Phys. France {\bf 50} (1989) 609.}
\nref\GDLP{E. Guitter, F. David, S. Leibler and L. Peliti, Phys. Rev. Lett. 
{\bf 61} (1988) 2949; J. Phys. (Paris) {\bf 50} (1989) 1787.}
\nref\KN{Y. Kantor and D. R. Nelson, Phys. Rev. Lett. {\bf 58} (1987) 2774
and Phys. Rev.  {\bf A 36} (1987) 4020.}
\nref\TAS{see also 
M. Baig, D. Espriu and J. Wheater, Nucl. Phys. {\bf B314} (1989) 587;
R. Renken and J. Kogut, Nucl. Phys. {\bf B342} (1990) 753; R. Harnish and
J. Wheater, Nucl. Phys. {\bf B350} (1991) 861.}
\nref\KJ{Y. Kantor and M.V. Jari\'c, Europhys. Lett. {\bf 11} (1990) 157.}
\nref\DIG{P. Di Francesco and E. Guitter, Europhys. Lett. {\bf 26} (1994)
455.}
\nref\DIGS{P. Di Francesco and E. Guitter, Phys. Rev. {\bf E50}(1994)4418.}
\nref\CGP{E. N. M. Cirillo, G. Gonnella, and A. Pelizzola, Phys. Rev. 
{\bf E53} (1996) 1479.}
\nref\BDIGG{M. Bowick, P. Di Francesco, O. Golinelli and E. Guitter, 
Nucl. Phys. {\bf B450} (1995) 463.}
\nref\CGPS{E. N. M. Cirillo, G. Gonnella, and A. Pelizzola, Phys. Rev.
{\bf E53} (1996) 3253.}
\nref\LG{see also 
A. E. Lobkovsky, S. Gentges, Hao Li, D. Morse and T. A. 
Witten, Science {\bf 270} (1995) 1482; E. M. Kramer and 
A. E. Lobkovsky, Phys. Rev. {\bf E53} (1996) 1465; 
A. E. Lobkovsky, Phys. Rev. {\bf E53} (1996) 3750.}
\nref\MA{D. C. Mattis, Phys. Lett. {\bf A56} (1976) 421, [Erratum, {\bf A60} (1977) 492].}
\nref\K{R. Kikuchi, J. Chem. Phys. {\bf 60} (1974) 1071.}
\nref\MO{T. Morita, Prog. Theor. Phys. {\bf 103} (1984) 103.}
\nref\BMP{C. Buzano, A. Maritan and A. Pelizzola, J. Phys. Cond. Matt. {\bf 6}(1994) 327.}
\nref\EA{S. F. Edwards and P. W. Anderson, J. Phys. {\bf F}: Meth. Phys. 5 (1975) 965.}
\nref\An{G. An, J. Stat. Phys. {\bf 52} (1988)  727.}
\nref\MOS{T. Morita, J. Stat. Phys. {\bf 59} (1990) 819.}
\nref\BMPS{D. Bensimon, D. Mukamel and L. Peliti, Europhys. Lett. {\bf 18} (1992) 269.}
\nref\ACB{R. Attal, S. Chaieb and D. Bensimon, Phys. Rev. {\bf E48} (1993) 2232.} 

\newsec{Introduction}

The statistical properties of polymerized membranes, or tethered surfaces have been 
widely discussed in the past few years [\xref\NPW-\xref\P]. 
A polymerized membrane is the two-dimensional 
generalization of a linear polymer [\xref\NPW-\xref\KKN]. 
Its energy involves both an in-plane elastic (strain)
contribution and an out-of-plane (bending) one. At low temperature, such a membrane 
with bending rigidity is asymptotically flat and its radius of gyration $R_{G}$ increases 
as the linear internal dimension $L$ of the surface [\xref\NP-\xref\GDLP]. 
As a function of temperature, 
the membrane without self-avoiding interaction (phantom membrane) undergoes a 
crumpling transition from the low temperature flat phase to a high temperature 
crumpled phase $(R_{G}\sim \sqrt{\ln L})$ [\xref\KN-\xref\TAS]. The mechanism of the transition, 
in particular the stability of the flat phase, is rather subtle and relies on the
coupling between in-plane and out-of-plane deformation modes \NP. 

Very generally, 2-dimensional membranes can be discretized into triangulations, whose faces
are endowed with natural Heisenberg spin variables, representing the direction of the local 
normal vector to the surface \NPW.
Polymerized membranes, which have a fixed connectivity, then translate into
statistical spin systems on the {\it regular} triangular lattice. The corresponding 
spin system is however involved, because the resulting spins are not independent 
variables. The  
constraint of being normal vectors to a surface causes a long range 
interaction between the spins, which
 stabilizes the ordered flat phase, as opposed to the case of the usual 
2-dimensional unconstrained Heisenberg model, always disordered [\xref\NPW,\xref\NP].    

The above mechanism clearly indicates the subtlety of the correspondence between 
geometrical objects and spin systems, especially in two dimensions. 
In order to understand this, 
several simple models have been	proposed. The simplest one is probably a square \
lattice model introduced by David and Guitter \DG.
The model is a discrete rigid bond square lattice, which is allowed  to fold onto itself
along its bonds in a two-dimensional embedding space. The constraint there makes
the folds propagate along straight lines.
In the presence of bending rigidity, the system is always flat.

A triangular lattice version of this 2D folding model was then introduced by
Kantor and Jari\'c \KJ.
Describing the (up or down) normal to each triangle by a spin variable $\sigma=\pm 1$, the model 
Hamiltonian translates into that of the Ising model with however constrains 
on the spin variables.
This results in several new features, and a totally new phase diagram, as compared with the 
usual Ising model [\xref\KJ-\xref\CGP]. 
Finally, the folding of the triangular lattice embedded in a 3-dimensional discrete space
has been formulated as a 96-vertex model [\xref\BDIGG,\xref\CGPS]. 

Constraints do not appear only in the physical degrees of freedom, like local
normal vectors. If the membrane has disorder, the disorder itself
can also be constrained in some cases.
For example, if one folds a piece of paper and makes a crease, this will generate a 
spontaneous curvature along the crease. If one now crumples the paper randomly by 
hand [\xref\KKN,\xref\LG], 
the generated spontaneous curvature will be directly related to the 
configuration of the normal vectors of the resulting crumpled configuration.
For the latter to be accepted as a physical configuration, the normal vectors 
should also obey some constraints. The induced disorder (in this case the induced random 
spontaneous curvature) should thus obey similar ``physical" constraints.

Here we study a simple model of folding with such a ``physical'' quenched randomness. 
As a model, we use the triangular lattice with quenched random bending rigidity 
and clarify the importance of the ``physical" constraints on the disorder.

The paper is organized as follows. In section 2, we first discuss the general
folding problem of the triangular lattice and recall some known facts about it.
We then describe the precise type of randomness
which we consider on the lattice and which we write as a Mattis-like 
spin system \MA\ with constraints. 
In section 3, we describe the cluster variation method that we shall apply to the study
of the thermodynamics of the system [\xref\K-\xref\BMP]. 
We first describe the procedure for a general disordered system. 
We then restrict ourselves to the 
hexagonal approximation in which the clusters are made of a maximum of six triangles.
Next,  we analyze the pure (without disorder) system again, 
as a particular limiting case with trivial disorder. 
The results for the fully disordered system are
presented in section 4. After giving a few results following from a reduced variable 
analysis, we present the complete  phase diagram of the system.
In section 5, we study other variants of the disorder. In particular, we 
discuss the specificity of the ``physical" constraint on the random bending rigidity.
We present for each type of disorder the corresponding phase diagram.
Some concluding remarks are gathered in section 6.

\newsec{Folding with Quenched Random Bending Rigidity}

In this section, we first recall the rules of folding for the triangular
lattice \KJ. We then introduce disorder in the problem in terms
of a {\it random bending rigidity}.

\subsec{Folding of the triangular lattice: pure case}

Let us consider a regular triangular lattice which can be folded
onto itself along its bonds. We allow only for complete foldings 
which result in {\it two-dimensional} folded configurations.
Each bond thus serves as a hinge between its two 
neighboring triangles and is in either one of the two 
states: folded (with the two neighboring triangles face to 
face) or not (with the two neighboring triangles side by side).  
A folded state of the system is entirely determined by the list 
of its folded bonds.
In this definition, the folding process may cause self-intersections
 and the model corresponds to a ``phantom'' membrane.
Also this does not distinguish between the different ways of folding which 
result in the same folded state.

%
\fig{The eleven local fold environments for a vertex. 
Folds are represented by
 thick lines. One of the two (opposite) possible spin 
configurations on the triangles
is also indicated.}
{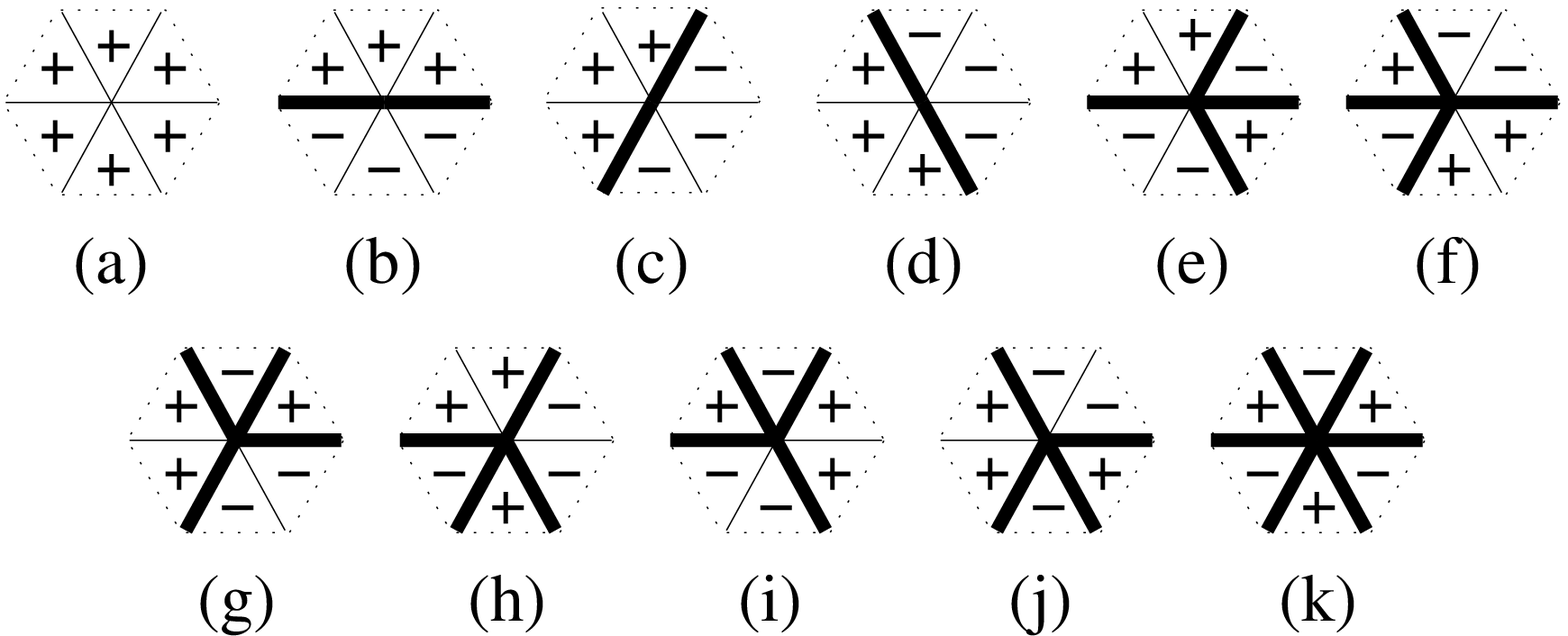}{10. truecm}
\figlabel\config
%

One can easily see that, among the $2^{6}$ possible fold 
configurations for the six bonds surrounding a  given vertex, 
only 11 states are allowed, corresponding to actual foldings 
of the surrounding hexagon \KJ. These configurations are displayed in
fig \config .
It can be checked that imposing everywhere one of these eleven
local environments is sufficient to define the folding 
consistently throughout the lattice.
The folding of the triangular lattice is thus simply expressed as
an 11-vertex model on the lattice.
Note that, even if the folding is defined locally, 
its nature is highly non-local.
Since all vertices in Fig.\config\  have an even number of 
elementary folds, these folds form folding lines 
without endpoints. Moreover all ``folded" vertices (vertices
(b)-(k) in Fig.\config ) have at least one fold on the left half
of the hexagon and one on the right. Thus folds are forced to
propagate through the entire lattice \DIG.

Folding can also be expressed with spin variables
$\sigma_{i}=\pm 1$ living on the elementary triangles, which 
indicate whether the $i$-th triangle faces up or
down in the folded state. The spin variable changes its sign 
between two neighboring triangles, if and only if their common
bond is folded, i.e. folding lines are domain walls of the 
spin system. One can think of the spin as the normal 
vector to the triangle. We depict the corresponding spin 
configurations on Fig.\config. 
Note that there are two spin 
configurations for each folded state, due to the degeneracy 
under reversal of all spins. 

Clearly, the only allowed vertex 
environments are those with exactly $0,3$ or $6$ 
surrounding up spins. 
In other words, for a spin configuration to correspond to a 
folded state, the six spins $\sigma_{i} (i=1,2,\cdots,6)$
around any vertex $v$ must satisfy the {\it local constraint} \DIG
\eqn\locons{\Sigma_v \equiv \sum_{i\ {\rm around} \ v} \sigma_i 
\ =\  0 \ {\rm mod} \ 3\ ,}
since $\Sigma_v=2(\hbox{number of up spins}) -6$ is a multiple of $3$
iff the number of up spins itself is a multiple of $3$.
Folding is thus expressed here as a {\it constrained} $Z_2$ 
{\it spin system}.

The statistical behavior of this system has been extensively 
studied, using a transfer matrix formalism [\xref\KJ,\xref\DIGS], 
the correspondence with a solvable three-coloring model \DIG\  
and a cluster variation method \CGP. 
Introducing a bending energy term 
$-J\sigma_{i}\sigma_{j}$ between nearest neighbors and a magnetic field term
 $-H\sigma_{i}$, the following model Hamiltonian was considered,
\eqn\hamising{ {\cal H}_{\rm Ising} = 
-J \sum_{(ij)} \sigma_i \sigma_j - H \sum_i \sigma_i \ . }
This is nothing but the Ising Hamiltonian, which is here however
coupled to the local constraint \locons .
In the folding context, the magnetization simply measures the projected
area of the lattice (i.e. the algebraic area of the domain enclosed by its boundary) 
and the magnetic field
can thus be interpreted as a lateral tension term.
For convenience we will use the
reduced coupling and magnetic field
\eqn\conv{K \equiv {J/k_BT} \ \ ; \ \  h \equiv H/k_BT \ .}

\nobreak
%
\fig{Phase diagram in the $(K,h)$ plane. 
Three first order lines $h=h_c(K),-h_c(K)$ ($K<K_c$) and $h=0$
($K>K_c$) separate the three phases $M=0$, $\pm 1$ and meet at the triple 
point $(K_c,0)$. The dashed line
represents the transition line between the disordered folded phase 
$M=0,M_{\rm st}=0$ and the compactly ordered folded phase 
$M=0, M_{\rm st}\neq 0$.}{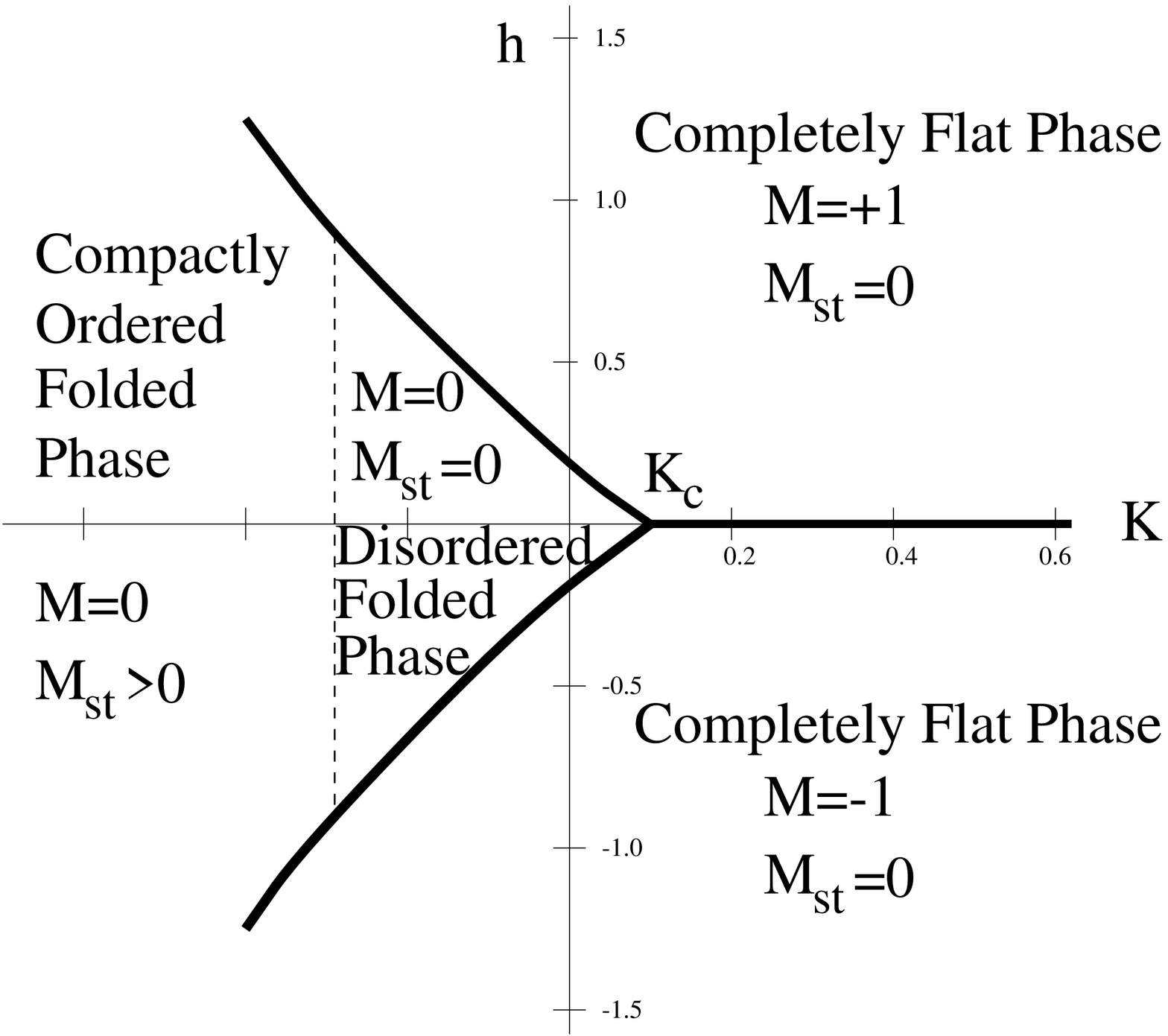}{7.cm}
\figlabel\phases
%
\nobreak

 The phase diagram of the system is shown in Fig.\phases. 
In order to characterize each phase, two order parameters are introduced.
One is the usual magnetization,
\eqn\magnet{M={1 \over N_t}\langle \big( \sum_{\bigtriangleup} \sigma_i +
\sum_{\bigtriangledown} \sigma_i \big) \rangle ,}
and the other is  the staggered magnetization 
\eqn\amagnet{M_{\rm st}={1 \over N_t}\langle \big( \sum_{\bigtriangleup} \sigma_i -
\sum_{\bigtriangledown} \sigma_i \big) \rangle ,}
where the sum is performed separately on triangles
pointing up and down, dividing the original triangular lattice into
two inter-penetrating sublattices. Here $N_t$ is the
total number of triangles in the system.
Three phases exist: a completely flat phase $(M=\pm 1 , M_{\rm st}=0)$, 
a disordered folded state $(M=0,M_{\rm st}=0)$ and a compactly ordered 
folded state $(M=0,M_{\rm st}\neq 0)$ [\xref\DIGS,\xref\CGP]. 
Note that the flat phase has a maximal magnetization $\vert M\vert =1$
and is indeed frozen in the pure completely flat state with all spins
aligned. There is no flat phase with intermediate ($0<\vert M\vert <1$)
magnetization.
Three first order transition lines $h=h_{c}(K),
-h_{c}(K)$ $(K<K_{c})$, and $h=0$ $(K>K_{c})$ separate the three phases 
$M=0,\pm 1$, with a triple point at $K_{c}\sim 0.1$ (estimated 
from either the transfer matrix \DIGS\ or the cluster variation \CGP\ approach).
For negative $K$, the transition 
between the disordered folded phase and a compactly ordered (antiferromagnetic) 
folded phase with staggered order parameter $M_{\rm st}\neq 0$ is found
to be continuous at $h=0$ \CGP. This transition 
is represented by the broken line of Fig.\phases, which intersects the 
horizontal ($h=0$) axis at $K\sim -0.284$.
We see here that, when compared with the usual unconstrained 
Ising model, the phase diagram has been strongly modified. 
In particular, it is now asymmetric with respect to $K$ with the
usual continuous ferromagnetic transition replaced by an abrupt
first order transition to a completely ordered phase.
\medskip

\subsec{Quenched random bending rigidity}

What kind of disorder should one introduce in the above 
model (Eq.\hamising)? Since we are dealing with folding of a
phantom object with a two-dimensional resulting folded state, 
we cannot distinguish between $\pm 180$ degree folds, thus 
we cannot introduce any spontaneous curvature term. 
Disorder will appear here in the form of a random bending
rigidity $K_{ij}$. The most general Hamiltonian with random 
nearest neighbor coupling $K_{ij}$ and a constant external field $h$ is
(we drop the $k_BT$ factor, thus using reduced coupling constants)
\eqn\hgeneral{ {\cal H}_{\rm random} = 
\sum_{(ij)}-K_{ij} \sigma_i \sigma_j - h \sum_i \sigma_i \ . }
There are several possible choices for $K_{ij}$. One possibility is the 
Edwards-Anderson model for spin-glasses \EA, where  
$K_{ij}=\pm K$ independently on each bond.
Such a model is interesting as it may have a spin glass phase.
In such a phase, spins are randomly oriented ($M=0$ and $M_{\rm st}=0$),
but the following order parameter $q$ takes a nonzero value 
\eqn\spgl{q={1 \over N_t}\sum_{i} \overline{\langle \sigma_i \rangle^{2}} .}
Here the upper bar stands for the quenched average over the 
randomness. We shall discuss this type of disorder in section 5.

As has been discussed in the introduction, 
we are interested in another type of ``physical" disorder
where the random bending rigidity has been generated by a prior
irreversible folding of the lattice [\xref\KKN,\xref\LG]. We have in mind the
picture of a crumpled piece of paper marked with irreversible 
creases.  
The effect of this irreversible crumpling is to impose the corresponding
crumpled state as the new ground state of the system. 
No frustration will occur as long as one does 
not strain the paper. The model Hamiltonian should thus have that 
crumpled state as its ground state and this information should be 
included in $K_{ij}$. One natural choice for $K_{ij}$ with a 
``random'' ordered phase and  no frustration is that of a Mattis Model \MA.  
In this model, the bending rigidities $K_{ij}$ are functions of a 
set of random face variables $\tau_{i}$:
\eqn\bemattis{ K_{ij}=K \tau_{i} \tau_{j}.}
The $\tau_{i}=\pm 1$ are ``frozen'' according to a specified 
probability distribution 
$\rho_{\tau}(\tau_{1},\tau_{2} \cdots,\tau_{N_t})$ reminiscent
of the first irreversible crumpling process.
The total Hamiltonian is then given by
\eqn\hammattis{ {\cal H}_{\rm Mattis} = 
-K \sum_{(ij)} \tau_{i}\tau_{j}\sigma_i \sigma_j - h \sum_i \sigma_i \ . }
As is well-known, if there is no constraint on spin variables, 
the following gauge transformation
\eqn\gauge{ \sigma_{i}' \rightarrow \sigma_{i}\tau_{i}}
makes the above model \hammattis\ in zero external field $h=0$ 
equivalent to the pure system \MA.
At $K=\infty$, the state with $\sigma_{i}=\tau_{i}$
 is recovered as the ground state. 

 In our model, the spin variables are constrained by
Eq.\locons\ and the above gauge transformation cannot be performed
since it does not preserve the folding constraint.
Moreover, the ground state $\sigma_i=\tau_i$ can be reached
only if the $\tau$ variables themselves obey the folding constraint:
\eqn\tlocons{\sum_{i\ {\rm around} \ v} \tau_i 
\ =\  0 \ {\rm mod} \ 3\ .}
 This type of $\tau$-configuration is what we call 
a ``physical" disorder. We shall restrict ourselves
 to this type of disorder in the following sections 3 and 4. We
 will return to other types of disorder (such as the Edwards-Anderson model)
 later in section 5.

{}From the above discussion, we understand the physical origin of 
the local constraint on the $\tau$-variables. Then what probability 
distribution $\rho_{\tau}(\tau_{1},\tau_{2} \cdots,\tau_{N_t})$ 
should we use for them?  
Since the $\tau$ variables obey the same constraint as the pure
$\sigma$ system described in the first part
of this section, we can take advantage of the solution of this
pure system by simply assuming that the $\tau$ distribution
is described by a particular (appropriately chosen) point in the  
disordered phase of the phase diagram of Fig.\phases.
A natural choice is the point at the origin of the phase diagram
($K=h=0$) since it then does not involve any energy parameter
for the disorder and treats as equiprobable all allowed	configurations
of ``physical" disorder.
In other words, we shall take for the
distribution $\rho_{\tau}(\tau_{1},\tau_{2} \cdots,\tau_{N_t})$ 
the density of the pure constrained problem at $K=h=0$.
Due to the constraint, this density remains non-trivial.

\newsec{Cluster Variation Method for the Disordered System}

In this section we explain in detail the hexagon 
approximation of the cluster 
variation method (CVM) generalized to a random system [\xref\K-\xref\BMP].
In subsection 3.1, we explain the method in the general case. 
In subsection 3.2, we apply the CVM to the pure system $(K_{ij}=K)$
as a limiting case with trivial disorder. As mentioned
above, this is also instrumental to give an explicit form for $\rho_{\tau}$
which is necessary to tackle the fully disordered case. 
We also discuss several symmetry breakings of the model.

\subsec{The CVM and its hexagon approximation}

The CVM is a closed-form approximation based on the 
minimization of an approximated free energy density functional,
 which is obtained by a truncation of the cluster expansion of the 
full free energy density functional appearing in the exact variational 
formulation of the problem [\xref\An,\xref\MOS].

Consider our spin system $\sigma_{i}$ with $N_t$ sites and 
Hamiltonian \hgeneral \BMP. The configuration of the random 
bond couplings $K_{ij}$ is specified by a probability distribution $P(\{K_{ij}\})$.
In our case of ``physical" disorder where $K_{ij}=K\tau_i\tau_j$, we shall
have:
\eqn\norp{P(\{K_{ij}=K\tau_i\tau_j\})=\rho_\tau(\tau_1,\cdots,\tau_{N_t}) .}
The discussion below is however more general.

In terms of a density matrix $\rho(\sigma_{1},\sigma_{2},\cdots,
\sigma_{N_t}|\{K_{ij}\})$ for each configuration $\{K_{ij}\}$, we
define the variational free energy associated with the Hamiltonian 
${\cal H}_{\rm random}(\sigma,\{K_{ij}\})$ (from now on, we use
the notation $\sigma = \{\sigma_i\}$) as
\eqn\free{{\cal F}(\{K_{ij}\})=\Big[\sum_{\sigma}\rho
(\sigma|\{K_{ij}\})[{\cal H}_{\rm random}(\sigma,\{K_{ij}\})
+\ln \rho(\sigma|\{K_{ij}\})]\Big]_{\rm min}  .}
The subscript min means that the above expression must be 
taken at its minimum with respect to $\rho(\sigma|\{K_{ij}\})$.
This is the well-known variational principle.
The minimization is performed at fixed $\{K_{ij}\}$ with the 
normalization constraint:
\eqn\norsig{\sum_\sigma \rho(\sigma|\{K_{ij}\})=1.}
The quenched free energy ${\cal F}$ is then given by
\eqn\qfree{{\cal F}=\sum_{\{K_{ij}\}}P(\{K_{ij}\}){\cal F}(\{K_{ij}\}),}
where the sum extends over all possible realizations of the disorder.
Upon introducing the generalized density matrix
\eqn\grho{\rho(\sigma,\{K_{ij}\})=P(\{K_{ij}\})\rho(\sigma|\{K_{ij}\}),}
one can easily show that
\eqn\qqfree{{\cal F}=\Big[ \sum_{\sigma,\{K_{ij}\}}\rho
(\sigma,\{K_{ij}\})[{\cal H}_{\rm random}(\sigma,\{K_{ij}\})
+\ln \rho(\sigma,\{K_{ij}\})]\Big]_{\rm min}+S_{\rm Dis}  }
where the minimization is now on a density $\rho(\sigma,\{K_{ij}\})$
for both $\sigma$ and $\{K_{ij}\}$ 
with the constraint
\eqn\pcons{\sum_{\sigma}\rho(\sigma,\{K_{ij}\})=P(\{K_{ij}\}).}
The quantity $S_{\rm Dis}$ is a constant term depending only on 
the probablity distribution for the disorder and reads
\eqn\disentropy{S_{\rm Dis}=-\sum_{\{K_{ij}\}}P(\{K_{ij}\})\ln P(\{K_{ij}\}).}
In such a scheme it can be shown that the quenched average 
$\overline{\phantom{AAA}}$ of the expectation value $<A(\sigma,\{K_{ij}\})>$ of an 
operator $A(\sigma,\{K_{ij}\})$ is given by
\eqn\expect{\eqalign{\overline{<A(\sigma,\{K_{ij}\})>}&
=\sum_{\{K_{ij}\}}P(\{K_{ij}\})<A(\sigma,\{K_{ij}\})>\cr 
&= \sum_{\{K_{ij}\},\sigma}\rho_{\rm min}(\sigma,\{K_{ij}\})A(\sigma,\{K_{ij}\}) .}}
where $\rho_{\rm min}(\sigma,\{K_{ij}\})$ is the density at the minimum of the
quenched free energy functional \qqfree.

The CVM is obtained by taking the thermodynamic limit $N_t\to \infty$ and 
truncating the cumulant expansion for the entropy ${\cal S}
\equiv - \sum_{\sigma,\{K_{ij}\}} \rho(\sigma,\{K_{ij}\})$ $\ln 
\rho(\sigma,\{K_{ij}\})$ appearing in \qqfree\ to a set of 
``maximal preserved clusters'' $\Gamma_{i},i=1,2,\cdots,r$
(and all their translated images). The variational principle will 
then be applied to the reduced density matrix 
$\rho_{\Gamma_{i}}(\sigma,\{K_{ij}\})$ associated with the maximal 
preserved clusters $\Gamma_{i}$, i.e. the minimization will be performed
on this reduced set of densities. 

\nobreak
%
\fig{Labeling of the spins on an elementary hexagon. Each site $i$ ($=1,\cdots,6$)
supports a spin variable 
$\sigma_{i}$ and a disorder variable $\tau_{i}$.
The reduction process from $\rho_{6}$ to $\rho_{2}$ and to $\rho_{1A(B)}$ is also indicated.
}{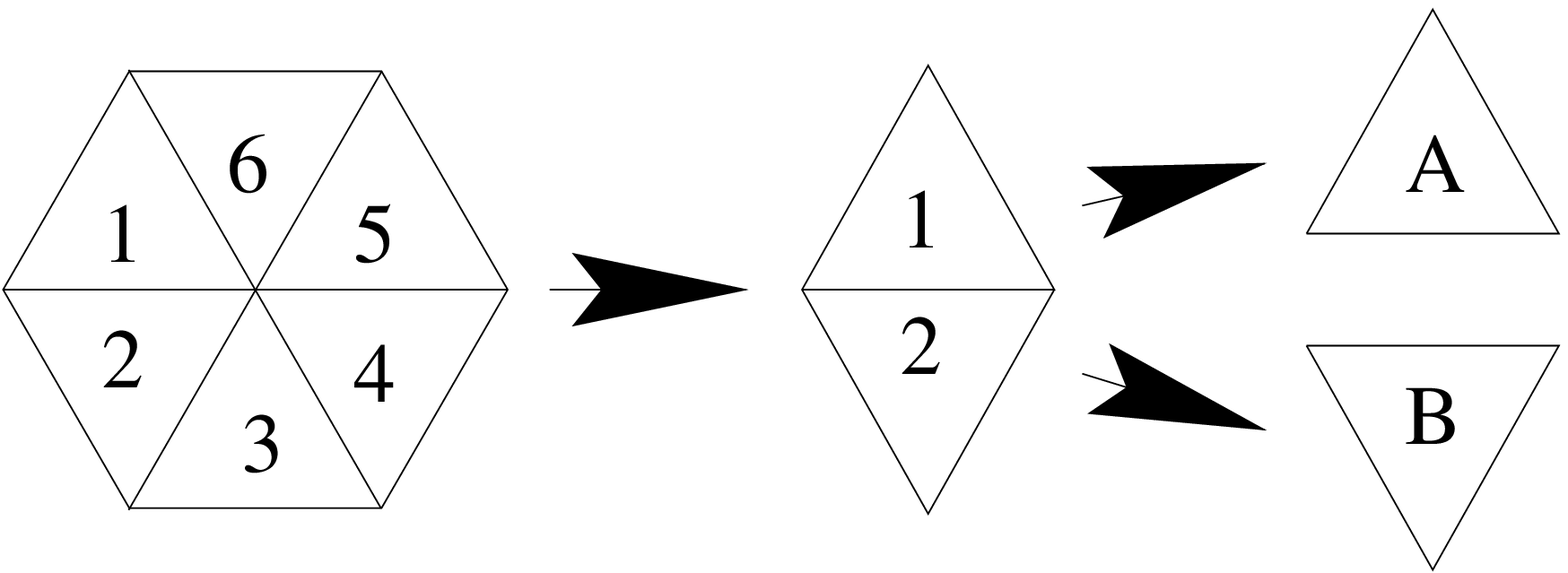}{7.cm}
\figlabel\hexagon
%
\nobreak

In the hexagon approximation for 
the triangular lattice, the largest clusters appearing 
in the expansion are hexagons. Hereafter we restrict our 
presentation to the case of the Mattis like 
coupling $K_{ij}=K\tau_{i}\tau_{j}$.
For the other cases, the generalization is straightforward.
We introduce the reduced density matrix for a hexagon.
\eqn\hex{\rho_{6}(\sigma_{1},\sigma_{2},\sigma_{3},\sigma_{4},\sigma_{5},\sigma_{6},
\tau_{1},\tau_{2},\tau_{3},\tau_{4},\tau_{5},\tau_{6})    .}
 The spins in the argument of $\rho_{6}$ follow each other counterclockwise 
in the hexagon, and the first one is on the A sublattice (see Fig.\hexagon)
i.e. is pointing up. 
This reduced density matrix represents the probability
for one hexagon to have fixed values of $\sigma$ and $\tau$. It is normalized
according to 
\eqn\norma{\sum_{\{\sigma \}}\rho_6(\{\sigma \},\{\tau \})=\rho_{\tau,6} (\{\tau \}),} 
 where $\rho_{\tau,6}$ is the 6-point probability for the disorder
 variable on a hexagon as obtained from the corresponding
 partial trace of $\rho_\tau(\tau_1,\cdots,\tau_{N_t})$ in the
 thermodynamic limit. We assume that this 6-point reduced density
 is the same for each hexagon, i.e. that the distribution of disorder is
 translationaly invariant.

 We also introduce the site and pair density matrices 
$\rho_{1A(B)}(\sigma_{1},\tau_{1})$, $\rho_{2}(\sigma_{1},\sigma_{2},\tau_{1},\tau_{2})$, 
which are defined as symmetrized partial traces of $\rho_{6}$ by
\eqn\trace{
\eqalign{
\rho_{2}(\sigma_{1},\sigma_{2},\tau_{1},\tau_{2})\equiv {1 \over 6} \sum_{{\sigma_{3},\sigma_{4},
\sigma_{5},\sigma_{6}}\atop {\tau_{3},\tau_{4},\tau_{5},\tau_{6}}} 
&[\rho_{6}(\sigma_{1},\sigma_{2},\sigma_{3},\sigma_{4},\sigma_{5},\sigma_{6},
\tau_{1},\tau_{2},\tau_{3},\tau_{4},\tau_{5},\tau_{6}) \cr
&+\rho_{6}(\sigma_{3},\sigma_{2},\sigma_{1},\sigma_{4},\sigma_{5},\sigma_{6},
\tau_{3},\tau_{2},\tau_{1},\tau_{4},\tau_{5},\tau_{6}) \cr
&+\rho_{6}(\sigma_{3},\sigma_{4},\sigma_{1},\sigma_{2},\sigma_{5},\sigma_{6},
\tau_{3},\tau_{4},\tau_{1},\tau_{2},\tau_{5},\tau_{6}) \cr
&+\rho_{6}(\sigma_{3},\sigma_{4},\sigma_{5},\sigma_{2},\sigma_{1},\sigma_{6},
\tau_{3},\tau_{4},\tau_{5},\tau_{2},\tau_{1},\tau_{6}) \cr
&+\rho_{6}(\sigma_{3},\sigma_{4},\sigma_{5},\sigma_{6},\sigma_{1},\sigma_{2},
\tau_{3},\tau_{4},\tau_{5},\tau_{6},\tau_{1},\tau_{2}) \cr
&+\rho_{6}(\sigma_{1},\sigma_{3},\sigma_{4},\sigma_{5},\sigma_{6},\sigma_{2},
\tau_{1},\tau_{3},\tau_{4},\tau_{5},\tau_{6},\tau_{2}) ], \cr
\rho_{1A}(\sigma_{1},\tau_{1})\equiv \sum_{\sigma_{2},\tau_{2}}
\rho_{2}(\sigma_{1},\sigma_{2}&,\tau_{1},\tau_{2})  ,\cr
\rho_{1B}(\sigma_{2},\tau_{2})\equiv \sum_{\sigma_{1},\tau_{1}}
\rho_{2}(\sigma_{1},\sigma_{2}&,\tau_{1},\tau_{2}).\cr
}}
Here we have introduced two site density matrices, $\rho_{1A}$ and 
$\rho_{1B}$, corresponding to the two inter-penetrating sublattices 
in which the triangular lattice can be divided (see Fig.\hexagon).

After the appropriate truncation of the cumulant expansion of $\cal S$
at the level of hexagonal clusters, we get
the approximate CVM quenched free energy per hexagon as a functional
of $\rho_6(\{\sigma_i\},\{\tau_i\})$ only (by implicit use of Eq.\trace )\An,
\eqn\fenergy{
\eqalign{&f(\rho_{6}(\{\sigma_{i}\},\{\tau_{i}\}))=-{3} K 
\Tr_{\sigma,\tau}(\tau_{1}\tau_{2}\sigma_{1}\sigma_{2}
\rho_{2}(\sigma_{l},\sigma_{2},\tau_{1},\tau_{2})) \cr
  &-h \Tr_{\sigma} (\sigma_{1}\rho_{1A}(\sigma_{1},\tau_{1})
   -h \Tr_{\sigma} (\sigma_{2}\rho_{1B}(\sigma_{2},\tau_{2})      \cr
&+\Tr_{\sigma,\tau}(\rho_{6}\ln \rho_{6})-3\Tr_{\sigma,\tau}(\rho_{2}\ln \rho_{2})
+\Tr_{\sigma,\tau}(\rho_{1A}\ln \rho_{1A})
+\Tr_{\sigma,\tau}(\rho_{1B}\ln \rho_{1B})                 \cr
&+\Tr_\tau\Big[\lambda_\tau(\{\tau_{i}\})(\Tr_{\sigma} 
\rho_{6}(\{\sigma_{i}\},\{\tau_{i}\})-\rho_{\tau,6}(\{\tau_{i}\}))\Big] \cr
&+S_\tau, \cr
}}
to be minimized with respect to $\rho_6(\{\sigma_i\},\{\tau_i\})$\foot{The cumulant expansion
of the entropy can be understood as follows. We write the truncated entropy as 
$${\cal S}= \sum_{\rm hexagons\ H} {\cal S}_H-\sum_{\rm pairs\ P} {\cal S}_P + \sum_{\rm triangles\ T} 
{\cal S}_T$$
where we must first subtract the contribution of pairs of neighboring triangles and 
re-add that of single triangles to avoid overcounting.
Noting that the numbers $N_H$, $N_P$, $N_{TA}$ and $N_{TB}$ of respectively
hexagons, pairs and triangles of the sublattices A and B satisfy $N_P/N_H=3$,
$N_{TA}/N_H=N_{TB}/N_H=1$, this leads to the entropy per hexagon appearing in \fenergy.}.
Here $\Tr$ stands for trace and $\lambda_\tau(\{\tau_{i}\})$ are Lagrange multipliers which 
ensure the normalization of $\rho_{6}(\{\sigma_i\},\{\tau_i\})$, according to Eq.\norma. 
$S_\tau$ is the entropy for 
the disorder variable $\tau_{i}$, for which we also use the CVM estimate:
\eqn\entropy{
S_\tau=\Tr_{\tau}(\rho_{\tau,6}\ln\rho_{\tau,6})-
3\Tr_{\tau}(\rho_{\tau,2}\ln \rho_{\tau,2})
+ 2 \Tr_{\tau}(\rho_{\tau,1}\ln \rho_{\tau,1})
,}
where $\rho_{\tau,2}$ and $\rho_{\tau,1}$ are 
partial (symmetrized) traces of $\rho_{\tau,6}$. 
With the above definitions, our free energy can be regarded as a 
function of $\rho_{6}$ only and taking 
the derivative with respect to a generic element 
of $\rho_{6}$ we find the stationarity conditions
\eqn\stationary{
\eqalign{\rho_{6}(\{\sigma_{i}\},\{\tau_{i}\})&=\exp[-\lambda_\tau(\{\tau_{i}\})+
{K \over 2}\sum_{i=1,6}
\tau_{i}\tau_{i+1}\sigma_{i}\sigma_{i+1}+{h \over 3}\sum_{i=1,6}\sigma_{i}]   \cr
&\times [\rho_{2}(\sigma_{1},\sigma_{6},\tau_{1},\tau_{6})\rho_{2}(\sigma_{1},\sigma_{2},\tau_{1},\tau_{2})
\rho_{2}(\sigma_{3},\sigma_{2},\tau_{3},\tau_{2}) \cr
&\ \ \rho_{2}(\sigma_{3},\sigma_{4},\tau_{3},\tau_{4})
\rho_{2}(\sigma_{5},\sigma_{4},\tau_{5},\tau_{4})
\rho_{2}(\sigma_{5},\sigma_{6},\tau_{5},\tau_{6})]^{1/2}                     \cr
&\times [\rho_{1A}(\sigma_{1})\rho_{1B}(\sigma_{2})\rho_{1A}(\sigma_{3})\rho_{1B}(\sigma_{4})
\rho_{1A}(\sigma_{5})\rho_{1B}(\sigma_{6})]^{-1/3}                        ,\cr
}}
with the convention $\sigma_7=\sigma_1$, $\tau_7=\tau_1$.

One can solve this set of equations with the definitions of Eqs.\trace\ 
and the normalization constraint \norma\ by the so-called natural 
iteration method \K. Starting from some assumption on $\rho_{2}$ 
and iterating the above equation, $\rho_{6}$ converges to a solution of
\stationary\ which is moreover a local {\it minimum} of the 
approximate free energy \fenergy  . To find a global minimum, it is in general necessary to start 
the iteration with 
different sets of initial conditions on $\rho_{2}$ appropriately chosen to reach 
the different expected phases.
At each step, the normalization condition \norma\ is recovered 
by adjusting the Lagrange
multiplier for each realization of the disorder on the hexagon.
Before going to the analysis of the model with disorder, in 
the next subsection we will revisit 
the pure case \CGP. One reason is that we need to fix the 
probability distribution for the disorder variables 
$\rho_{\tau}(\{\tau_{i}\})$.

\subsec{Pure case and explicit form for $\rho_{\tau}$}

In this section, we reconsider the pure case for $h=0$ in detail within the 
CVM approximation as a particular trivial realization of disorder
where all $K_{ij}$ are constant and equal to $K$ \CGP. We only need to consider 
the ``pure'' 6-point functions $\rho_{6} (\sigma_{i})$, 
which does not depend on the $\tau_{i}$ variable any longer.
We can easily recognize that the $2\times 11=22$ elements of $\rho_{6}$, 
which correspond to the weights for each state in Fig.\config, are not 
all independent since some of the states are related by simple symmetries 
and should thus have the same weight. 
This of course assumes that the corresponding symmetries are not spontaneously
broken.  Hereafter we only consider the system for $h=0$
and consider three types of solutions corresponding to the three
different symmetries of the spin system in the phase diagram of Fig.\phases\ \CGP:

\item{(1)} Disordered folded phase: we do not allow for any
spontaneous symmetry breaking in the system. Using rotational
symmetry and the symmetry under reversal of all spins, we end up 
with only 4 independent weights $Z_{0,1,2,3}$ corresponding
to vertices with respectively 0,2,4 or 6 surrounding folds.

\item{(2)} Ferromagnetic phase: we allow for a spontaneous 
ferromagnetic symmetry breaking ($M\neq 0$). Then the two vertices 
with no fold have different weight $Z_0$ and $\overline{Z_0}$
according to their $\pm 6 $ magnetization. The other vertices
are neutral in this respect and we end up with 5 different weights.

\item{(3)} Antiferromagnetic phase: we allow for a spontaneous 
antiferromagnetic symmetry breaking ($M_{\rm st}\neq 0$). 
Then all weights $Z_i$ have to be doubled into $(Z_i,\overline{Z_i})$
except for the vertex with no fold ($i=0$) which is neutral
in the staggered magnetization.  We end up with 7 different weights.

\bigskip
\noindent
%
$$\vbox{\offinterlineskip
\halign{\tv\quad # \quad & \tv \quad 
# \quad & \tv \quad # \quad & \tv \quad # \quad & \tv \quad # 
\quad \tv \cr 
\noalign{\hrule}
\tvi ${\rm Spin}\ {\rm Conf.}$ & ${\rm (1)\ Dis}.$ & ${\rm (2)\ Ferro.}$ & 
${\rm (3)\ A.Ferro.}$ & ${\rm Deg.}$ \cr
\noalign{\hrule}
\tvi  $++++++$ &  $Z_{0}$ &  $ Z_{0} $ & $ Z_{0} $  & $ 1 $   \cr
\tvi  $------$ &  $Z_{0}$ &  $ \overline{Z_{0}} $ & $ Z_{0} $  & $ 1 $   \cr
\tvi  $+++---$ &  $Z_{1}$ &  $ Z_{1} $  & $ Z_{1} $  & $ 3 $   \cr
\tvi  $---+++$ &  $Z_{1}$ &  $ Z_{1} $  & $ \overline{Z_{1}} $  & $ 3 $   \cr
\tvi  $+--++-$ &  $Z_{2}$ &  $ Z_{2} $  & $ Z_{2} $  & $ 6 $   \cr
\tvi  $-++--+$ &  $Z_{2}$ &  $ Z_{2} $  & $ \overline{Z_{2}} $  & $ 6 $   \cr
\tvi  $+-+-+-$ &  $Z_{3}$ &  $ Z_{3} $  & $ Z_{3} $  & $ 1 $   \cr
\tvi  $-+-+-+$ &  $Z_{3}$ &  $ Z_{3} $  & $ \overline{Z_{3}} $  & $ 1 $   \cr
\noalign{\hrule} }} $$
\nobreak
\centerline{{\bf Table I:} Independent hexagon spin configurations.}
\centerline{The corresponding elements of $\rho_{6}$ and the degeneracies are indicated.}
\vskip 1cm
%
\noindent

The spin configurations, their degeneracies and the notations for their weights 
are summarized in Table I.
Of course, the case (1) can be recovered from either case (2) or (3) as 
a particular realization with no spontaneous symmetry breaking (i.e 
$Z_i=\overline{Z_i}$ for all $i$). 
Also, we assume that the two (ferromagnetic and antiferromagnetic) symmetries
cannot be broken simultaneously.
We thus need to study only the cases (2) and (3) above to  
get the complete phase diagram of the system (here at $h=0$).

 At first we consider the ferromagnetic case (2).
In this case, the stationarity condition
 reduces to the following nonlinear equations between the weights
$Z_{0},\overline{Z_{0}},Z_{1},Z_{2},Z_{3}.$

\eqn\z{
\eqalign{
Z_{0}&=\exp(-\lambda+3K)(y_{++})^{3}/(y_{+})^{2}    \cr
\overline{Z_{0}}&=\exp(-\lambda+3K)(y_{--})^{3}/(y_{-})^{2}  \cr
Z_{1}&=\exp(-\lambda+K)(y_{++})(y_{--})(y_{+-})/(y_{+})(y_{-})  \cr
Z_{2}&=\exp(-\lambda-K)(y_{++})^{1/2}(y_{--})^{1/2}(y_{+-})^{2}/(y_{+})(y_{-})     \cr
Z_{3}&=\exp(-\lambda-3K)(y_{+-})^{3}/(y_{+})(y_{-})  ,\cr
}}
involving a single Lagrange multiplier $\lambda$.
Here, $y_{++},y_{+-},y_{--},y_{+}$ and $y_{-}$ are two- and one-point 
functions $\rho_2(\sigma_{1}\sigma_{2})$ and $\rho_1(\sigma_{1})$
(there is no difference here between sublattices A and B), which are defined 
as follows, 
\eqn\y{
\eqalign{
y_{++}&=\rho_2(++)=Z_{0}+2Z_{1}+2Z_{2}    \cr
y_{+-}&=\rho_2(+-)=\rho_2(-+) =Z_{1}+4Z_{2}+Z_{3}  \cr
y_{--}&=\rho_2(--) =\overline{Z_{0}}+2Z_{1}+2Z_{2}  \cr
y_{+}&=\rho_1(+)=Z_{0}+3Z_{1}+6Z_{2}+Z_{3}    \cr
y_{-}&=\rho_1(-)=\overline{Z_{0}}+3Z_{1}+6Z_{2}+Z_{3}       .\cr
}}
 The above equations imply the following simple relations:
\eqn\prelation{
{(Z_{0}\overline{Z_{0}})^{1/2} \over Z_{1}}
={Z_{1}\over Z_{2}}={Z_{2}\over Z_{3}}
=\exp(2K){(y_{++})^{1/2}(y_{--})^{1/2}   \over  y_{+-}  }                     .}
Introducing the two reduced variables
\eqn\ratio{
x\equiv {Z_1\over (Z_{0}\overline{Z_{0}})^{1/2}}, \phantom{AAAA} 
y\equiv {\overline{Z_{0}} \over Z_{0}}    .
}
we can express all the weights in terms of $x$, $y$ and 
the  normalization factor $w_0\equiv (Z_{0}\overline{Z_{0}})^{1/2} $.
\eqn\zxyw{\eqalign{
Z_{0}&=y^{+1/2} w_{0}, \phantom{AAAAA}  \overline{Z_{0}}=y^{-1/2} w_{0}    \cr
Z_{1}&=x w_{0},         \phantom{AAAAA}  Z_{2}=x^{2} w_{0} , \phantom{AAAAA} 
Z_{3}=x^{3} w_{0} . 
}}

The above equations \z\ reduce to the following 
nonlinear equations for the reduced variables,
\eqn\prely{ y=    \big({ y+2y^{1/2}(x+x^{2})  
\over 1+2y^{1/2}(x+x^{2}) } \big)^{3}    
     \big({    1+y^{1/2}(3x+6x^{2}+x^{3}) \over
   y+y^{1/2}(3x+6x^{2}+x^{3})  }\big )^{2}      .}
\eqn\prelx{  x={y(x+4x^{2}+x^{3})   \over 
u(1+2y^{1/2}(x+x^{2}))^{1/2}(y+2y^{1/2}(x+x^{2}))^{1/2}  }.}
where $u=\exp(2K)$. The parameter $y$ measures the 
spontaneous ferromagnetic symmetry breaking while the parameter $x$ measures 
the fugacity per folded bond.

We can easily see that Eq.\prely\ has two obvious solutions: a solution 
$y=1$ and $x$ arbitrary and a solution $x=0$ and $y$ arbitrary.
The latter solution is also a solution of Eq.\prelx. It means that each
vertex of the membrane can be only in one of
the two configurations without fold $Z_{0}$ or $\overline{Z_{0}}$. The solution cannot 
determine the ratio $y={\overline Z_0}/Z_0$, i.e the proportion of each state.
However, with only these two vertices at hand, no fold can be ever created
and the only possible global states for the lattice are the state with
all spins up ($M=1$) and that with all spins down ($M=-1$).
The above solution simply describes an arbitrary superposition 
of these two (symmetric) pure flat states. The fact that the membrane
is indeed frozen in a pure completely flat state is further confirmed by computing
the entropy which is found to be exactly zero, and by computing the
free energy, which is found to be $f=-3K$ per hexagon, as expected
(there are 3 bonds per hexagon).

The first solution with $y=1$ means that the spontaneous symmetry breaking 
does not occur and that the membrane is in the disordered folded state 
$(Z_{0}=\overline{Z_{0}})$.
The value of $x$ is then fixed by Eq.\prelx \CGP:
\eqn\solx{x={(2-u)+\sqrt{(3-u-u^2)}\over (2u-1)}}
which has a solution for $K\le \ln((1+\sqrt{13})/2)/2$.
Comparing the corresponding free energy to that of the pure
flat state, we get a first order transition from disordered folded to 
purely flat at $K_c\sim 0.1013$ \CGP.

We also looked numerically for another non-trivial solution with spontaneous symmetry 
breaking $(y\neq 1)$ and intermediate magnetization ($x\neq 0$) but did not find any. 
We conclude that there 
is no possible flat phase with $0<|M|<1$ and the above three phases 
$(M=\pm 1$ or $0)$ are the only stable ones for positive $K$.

\nobreak
%
\fig{Probability distribution for each disorder configuration. We also show their 
degeneracies}{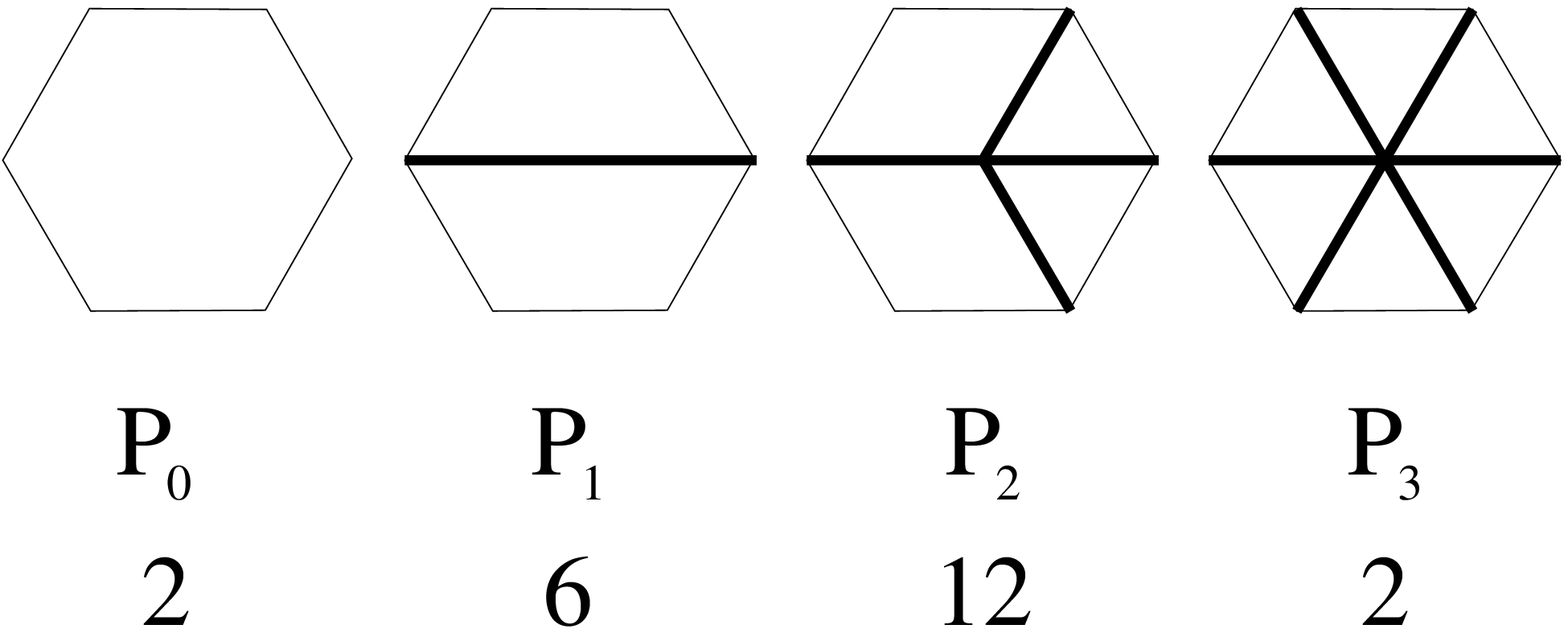}{7.cm}
\figlabel\probt
%
\nobreak

As has been discussed previously, the above analysis is also instrumental
for the estimation of the  
probability  distribution $\rho_{\tau,6}(\{\tau_i\})$ of the disorder variables $\tau$. 
We can use indeed for $\rho_{\tau,6}$ the distribution $\rho_6$
above at $K=h=0$, characterized by $y=1$ and $x=2$ \CGP. 
In other words, if we
define $P_{0,1,2,3}$ as the weights $\rho_{\tau,6}$ for the local
realizations of disorder with 0,2,4 or 6 creases around the vertex,
we learn that the ratios $P_1/P_0$, $P_2/P_1$ and $P_3/P_2$ must
all be identical and equal to 2.
Their values are then fixed by the normalization:
\eqn\normp{2P_0+6P_1+12P_2+2P_3=1}
leading to \CGP:
\eqn\valp{P_0={1\over 78}, P_1={2\over 78}, P_2={4\over 78}, P_3={8\over 78}.}

More generally, we can parametrize the distribution of $P$ with one parameter 
$\alpha$ equal to the ratios $\alpha=P_{1}/P_{0}=P_{2}/P_{1}=P_{3}/P_{2}$.
Beside the natural value $\alpha=2$ above, the limiting case $\alpha=0$ 
describes a membrane with no crease and $K_{ij}=K$ everywhere, while 
$\alpha=\infty$ describes a membrane with creases everywhere and $K_{ij}=-K$
on each bond. 

\medskip
Now we discuss the compactly ordered (antiferromagnetic) folded phase (3). 
There are 7 independent weights 
for $h=0$ (see Table I) and in this case it is convenient to use 
staggered variables $\eta_{i}=(-1)^{i-1} \sigma_{i}$ with
$(-1)^{i-1}=1$  on triangles belonging to the sublattice A
and $(-1)^{i-1}=-1$ on triangles belonging to the sublatice B.
The corresponding two-point function is simply
\eqn\twopeta{\rho_{\eta,2}(\eta_1,\eta_2)=\rho_2(\eta_1,-\eta_2).}
About the one-point function, we have the symmetry 
$\rho_{1A}(\sigma)=\rho_{1B}(-\sigma)$ in the antiferromagnetic phase,
leading to only one (A/B independent) one-point function for
$\eta$:
\eqn\onepeat{\rho_{\eta,1}(\eta)=\rho_{1A}(\eta)=\rho_{1B}(-\eta)}

As before, the solution of the non-linear stationarity equations for
the 7 weights can be parametrized as:
\eqn\zaf{
\eqalign{
Z_{0}&=w_{0}, \phantom{AAAAA} Z_{1}=x y^{1/2}w_{0}, \phantom{AA} 
\overline{Z_{1}}= x y^{-1/2} w_{0}  \cr
Z_{2}&=  x^{2} y^{1/2} w_{0},  \phantom{A} \overline{Z_{2}}= x^{2} y^{-1/2} w_{0} 
\phantom{A} Z_{3}= x^{3} y^{3/2}  w_{0}, \phantom{A}
\overline{Z_{3}}=  x^{3} y^{-3/2} w_{0} .\cr
}}
with two reduced variables $x$ and $y$ solutions of,
\eqn\ratioaf{
y={  y_{++} \over y_{--}} \big( {y_{-} \over y_{+}}\big)^{2/3},
\phantom{AAAAA}x=  u^{-1} {  y_{++}^{1/2} y_{--}^{1/2} \over y_{+-}}  .
}
where
\eqn\yaf{
\eqalign{
y_{++}&=\rho_{\eta,2}(++)=Z_{1}+3Z_{2}+\overline{Z_{2}}+Z_{3}    \cr
y_{+-}&=\rho_{\eta,2}(+-)=\rho_{\eta_2}(-+)=Z_{0}+Z_{1}+\overline{Z_{1}}+Z_{2}+\overline{Z_{2}}  \cr
y_{--}&=\rho_{\eta,2}(--)=\overline{Z_{1}}+3\overline{Z_{2}}+Z_{2}+\overline{Z_{3}}  \cr
y_{+}&=\rho_{\eta,1}(+)=Z_{0}+2Z_{1}+\overline{Z_{1}}+4Z_{2}+2\overline{Z_{2}}+Z_{3}    \cr
y_{-}&=\rho_{\eta,1}(-)=Z_{0}+2\overline{Z_{1}}+Z_{1}+4\overline{Z_{2}}+2Z_{2}+\overline{Z_{3}}       .\cr
}}
In the equations \ratioaf ,  the global normalization $w_0$ drops out, so 
the equations can be solved for $x$ and $y$ as functions of $u=\exp (2K)$.
Again the variable $x$ measures the fugacity for each fold and $y$ measures the 
antiferromagnetic spontaneous symmetry breaking.
Solving the above equations numerically by iteration, we find a continuous
transition from a disordered folded state ($y=1$) to a compactly folded
ordered phase ($y\neq 1$) at $K_{\rm st}=-0.2838$.
The value of $K_{\rm st}$ can be found simply by linearizing the equations
\ratioaf\ by writing $y=1+\epsilon$. This fixes the value of $x$ to be 
the real solution $x_{\rm st}$ of 
\eqn\tobefound{x^3-21x^2-12x-4=0}
that is
\eqn\xst{x_{\rm st}= 7 +(387 +2 \sqrt{223})^{1/3} +53/(387 + 2 \sqrt{223})^{1/3}}
and $u$ to be
\eqn\inphil{u_{\rm st}={1+4 x_{\rm st} +x_{\rm st}^2 \over 1+2x_{\rm st}+2x_{\rm st}^2}.}

\newsec{Results for the fully disordered system}

 In this section we analyze the fully disordered case (Eq.\hammattis) within
 the CVM approximation.
 In subsection 4.1, we study the system for $h=0$ for several
 values of the parameter $\alpha$ for the disorder weights. 
Next we fix $\alpha=2$ and proceed to the general $(K,h)$ case in subsection 4.2. 
We obtain the $(K,h)$-phase diagram by use of the natural iteration method.

\subsec{Analysis with reduced elements of $\rho_{6}$ ($h=0$ case)}

\noindent
%
\fig{The 38 independent local fold environments for each vertex for the 4 different local realizations of 
disorder (0,2,4,or 6 creases). Each disorder configuration is shown at the left 
hand side of each group. To their right, we show the spin configurations, the weights and the 
degeneracies.    
We have represented by
thick lines the domain walls for the gauged spin variable
$\eta_{i}=\sigma_{i}\tau_{i}$. A subscript $i,j$ indicates a configuration with $2i$
creases and $2j$ folds. The superscript $\pm$ indicates a $\pm$ contribution to $F$.
In a $F=0$ phase, equating the $+$ and $-$ weights leaves us with 22 independent weights.}
{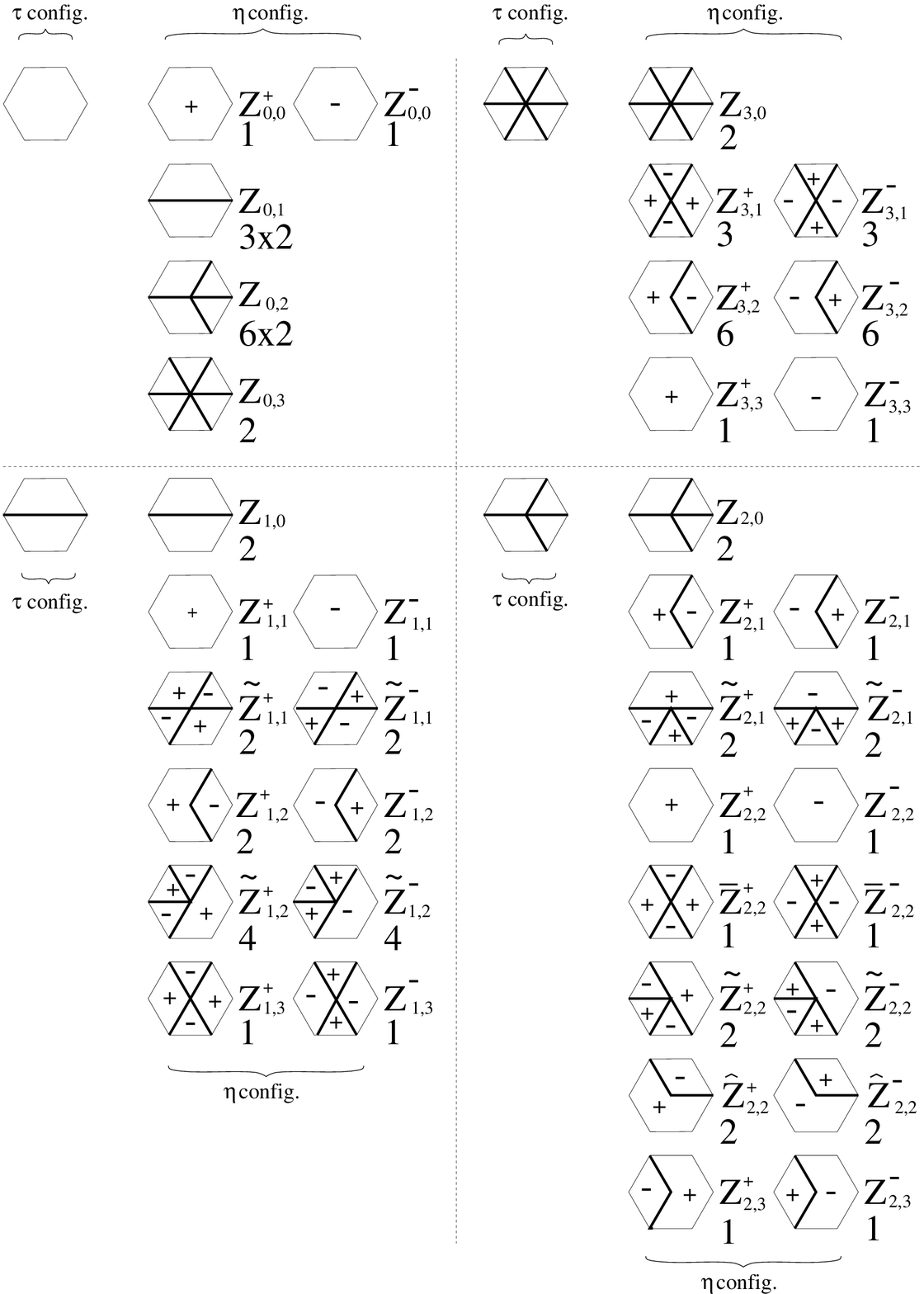}{9.5 truecm}
\figlabel\dconfig
%
\noindent

As in the previous section, we will study the fully disordered system 
for \break $h=0$ by reducing the number of elements of the 6-point density matrix. 
In the case with disorder, the symmetries of the elements of  
the 6-point function $\rho_6(\{\sigma_{i}\},\{\tau_{i}\})$ 
depend also on the symmetry of the disorder ($\tau$-variables), in addition to 
the symmetries of the spin variable $\{\sigma\}$ itself.

In the previous section, we have studied  the pure model with an 
antiferromagnetic  spontaneous symmetry 
breaking for $K<0$. There we have used 
the staggered variables as $\eta_{i}=(-1)^{i-1}\sigma_{i}$. 
If we regard the pure system for $K<0$ as
 a trivial disordered system with $\rho_{\tau,6}(1,-1,1,-1,1,-1)=1$,
 i.e. $\tau_i$ is fixed to $(-1)^{i-1}$, and $K>0$, 
the above staggered variables can be written as 
\eqn\nstagv{\eta_{i}=\sigma_{i}\tau_{i}.}
The motivation for introducing the staggered variables is that in these variables 
the antiferromagnetic order parameter $M_{\rm st}$ is simply written as
\eqn\mstag{M_{\rm st}={1 \over N_t}\langle \big( \sum_{i} {\eta_i} \big) \rangle ,}
 and we don't have to differentiate between the A- and B-sublattices.
That is, the staggered variables $\{\eta_{i}\}$ are more natural than the original 
variables $\{\sigma_{i}\}$ when one discusses the antiferromagnetic symmetric case.
  
In the fully disordered system, we are mainly interested in the spontaneous 
symmetry breaking at $K>0$ of the following ``frozen'' order parameter\foot{The
reader might wonder whether the $\tau\to -\tau$ symmetry could lead to a zero
quenched average of the order parameter. However, this symmetry implies only that
whenever $\rho_{\rm min}(\{\sigma\},\{\tau \})$ is a solution of the
variational equations, $\rho_{\rm min}(\{\sigma\},\{-\tau \})$ is also a solution,
but does not imply that $\rho_{\rm min}(\{\sigma\},\{\tau \})=\rho_{\rm min}(\{\sigma\},\{-\tau \})$.
}
\eqn\frozen{F={1 \over N_t}\langle \overline{\big( \sum_{i} {\sigma_i\tau_{i}} \big) \rangle}.}
This order parameter judges whether or not the membrane is trapped
in the randomly oriented phase, characterized by the disorder 
variables $\{\tau_{i}\}$.  As in the antiferromagnetic case, it is natural to use the 
following ``gauged'' variables,
\eqn\stagv{\eta_{i}=\sigma_{i}\tau_{i}.}

Using these new gauged variables, we classify the elements of 6-point functions 
by the symmetries of both the spin configuration and  the disorder configuration.
 Hereafter we only consider the system for $h=0$
and allow for two types of solutions  which correspond to whether the 
frozen order exists $(F\neq 0)$ or not $(F=0)$.

\item{(1)} Disordered folded phase: we do not allow for any
spontaneous symmetry breaking in the system $(M,M_{\rm st}$ and $F=0)$.
Each of the 4 elementary types of disorder configuration (with 0,2,4 or 6 creases) 
leaves us with a certain number of symmetries, including that
under reversal of all spins. We use these symmetries on the $\eta$ variables
to reduce the number of weights. 
We end up with only 22 independent weights in this case.

\item{(2)} Frozen phase: we allow for a spontaneous 
symmetry breaking of the frozen order parameter ($F\neq 0$).  
Then all weights have to be doubled except for those vertices which are neutral
in the gauged magnetization ($\sum_1^6 \eta_i=0$).  
We end up with 38 different weights in this case.

In Fig.\dconfig, we have summarized the results of this
symmetry analysis in the case (2). Note that case
(1) can always be seen as a particular case of case (2) with
extra symmetries.
On the left hand side of each group, we show the 
disorder configuration $\{\tau_{i}\}$.
To its right, we present the spin configurations $\{\eta_{i}\}$
which are independent from each other. We also indicate the notations 
for their weights and their degeneracies.
The two indices $i,j$ in $Z_{i,j}$ indicate a configuration
with $2i$ creases and $2j$ folds in the $\sigma$ variable.

\noindent
%
\fig{Definitions of two-point functions. 
Gauged spin variables ${\eta_{i}}=\sigma_{i}\tau_{i}$ are used.
On the left hand side, we show the disorder configuration $\tau$.}
{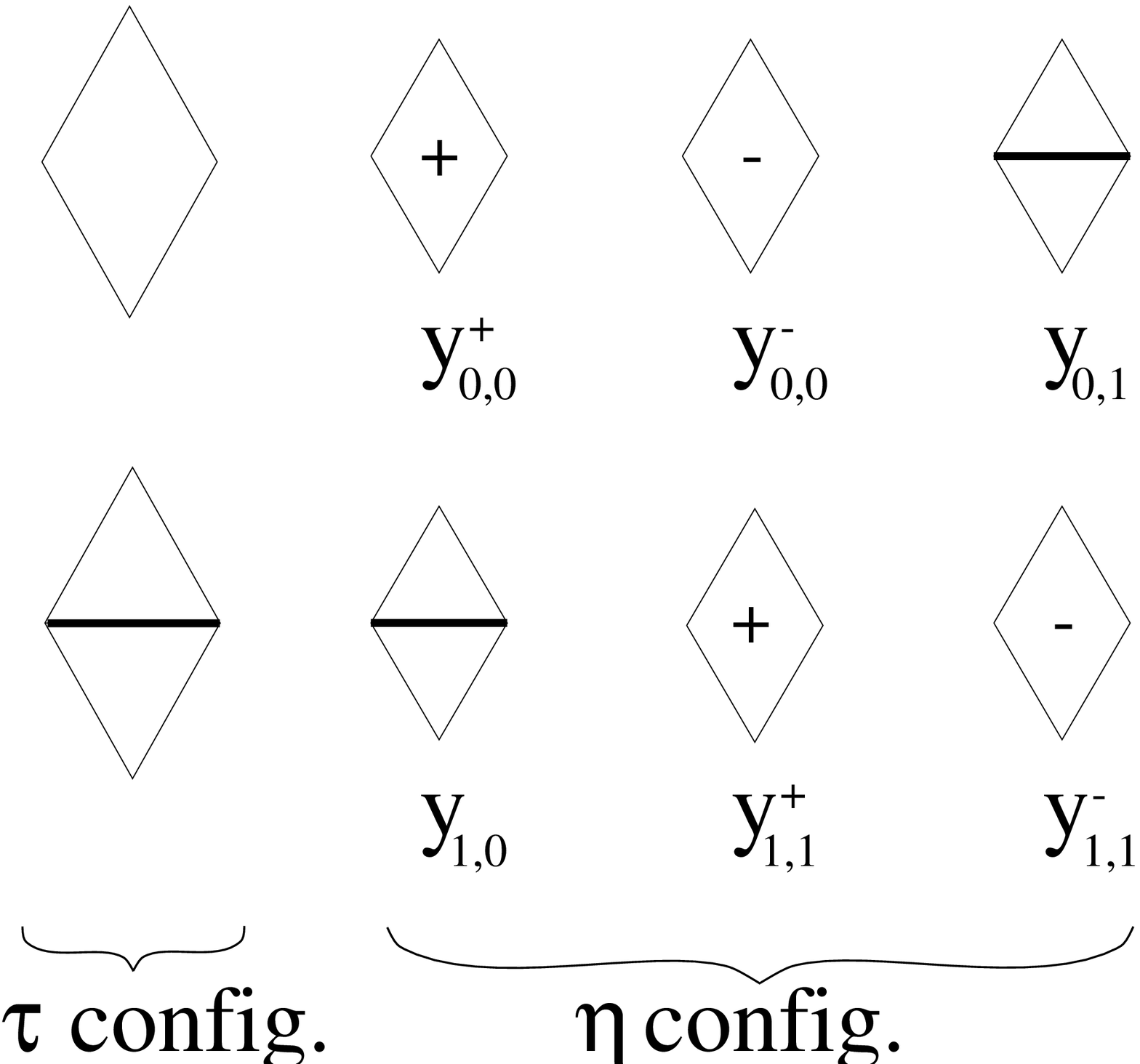}{6. truecm}
\figlabel\twopoint
%
\noindent

 In terms of these elements of the 6 point function, we define 
the two-point functions as follows,
\eqn\yy{
\eqalign{
y_{0,0}^{\pm}&=Z_{0,0}^{\pm}+2Z_{0,1}+2Z_{0,2}+2Z_{1,0}+2Z_{1,1}^{\pm }+
2{\tilde Z}_{1,1}^{\pm } +2Z_{1,2}^{\pm} \cr
&+2{\tilde Z}_{1,2}^{\pm}+2Z_{2,0}+2Z_{2,1}^{\pm}+2{\tilde Z}_{2,1}^{\pm}
+2Z_{2,2}^{\pm}+2{\hat Z}_{2,2}^{\pm},    \cr
y_{0,1}&=Z_{0,1}+4Z_{0,2}+Z_{0,3}+{\tilde Z}_{1,1}^{+} +{\tilde Z}_{1,1}^{-}
+Z_{1,2}^{+}+Z_{1,2}^{-}  \cr
&+3{\tilde Z}_{1,2}^{+}+3{\tilde Z}_{1,2}^{-}+Z_{1,3}^{+}+Z_{1,3}^{-}
+{\tilde Z}_{2,1}^{+}   \cr
&+{\tilde Z}_{2,1}^{-}+{\overline Z}_{2,2}^{+}+{\overline Z}_{2,2}^{-}+2{\tilde Z}_{2,2}^{+}+
2{\tilde Z}_{2,2}^{+}   \cr
&+{\hat Z}_{2,2}^{+} + {\hat Z}_{2,2}^{-} + Z_{2,3}^{+} + Z_{2,3}^{-} , \cr
y_{1,0}&=y_{0,1}(Z_{i,j}\to Z_{j,i}), \cr
y_{1,1}^{\pm}&=Z_{1,1}^{\pm}+Z_{1,2}^{\pm}+Z_{1,2}^{\mp} +2{\tilde Z}_{1,2}^{\pm}
+Z_{1,3}^{\pm}+Z_{2,1}^{\pm}+Z_{2,1}^{\mp}  \cr
&+4Z_{2,2}^{\pm}+2{\overline Z}_{2,2}^{\pm}+4{\hat Z}_{2,2}^{\pm}+2{\hat Z}_{2,2}^{\mp}
+4{\tilde Z}_{2,2}^{\pm}   \cr
&+3Z_{2,3}^{\pm}+Z_{2,3}^{\mp}+2{\tilde Z}_{2,1}^{\pm}+Z_{3,3}^{\pm}+3Z_{3,2}^{\pm}
\cr &+Z_{3,2}^{\mp} + Z_{3,1}^{\pm}  .   \cr
}}
The upper right index of $y$ means that the gauged spin 
configuration $\{\eta_{i}\}$
have the corresponding positive (or negative)
contribution to $F$. Again, the first lower right index indicate whether there is a crease 
line (1) or not (0). The second index means that there is a fold (1) or not (0) 
in the original $\sigma$ variables. Based on these two-point functions, 
we also introduce the following one point functions, 
\eqn\y{
y^{\pm}=y_{0,0}^{\pm}+y_{0,1}+y_{1,0}+y_{1,1}^{\pm}. }

In terms of these functions, we write down the stationarity conditions. For example,
 let us show those for $Z_{0,0}^{\pm},Z_{0,1},Z_{0,2}$ and $Z_{0,3}$.
\eqn\dstation{
\eqalign{
Z_{0,0}^{+}&=\exp(-\lambda_{0}+3K)(y_{0,0}^{+})^{3}/(y^{+})^{2},    \cr
Z_{0,0}^{-}&=\exp(-\lambda_{0}+3K)(y_{0,0}^{-})^{3}/(y^{-})^{2},    \cr
Z_{0,1}&=\exp(-\lambda_{0}+K)(y_{0,0}^{+})(y_{0,0}^{-})(y_{0,1})/(y^{+})(y^{-}),  \cr
Z_{0,2}&=\exp(-\lambda_{0}-K)(y_{0,0}^{+})^{1/2}(y_{0,0}^{-})^{1/2}(y_{0,1})^{2}/(y^{+})(y^{-}),     \cr
Z_{0,3}&=\exp(-\lambda_{0}-3K)(y_{0,1})^{3}/(y^{+})(y^{-})  .
}}
Here, due to the above symmetries, we need only 4 Lagrange multipliers 
$\lambda_{0,1,2,3}$, one for each of the 4 elementary types of disorder in Fig.\probt. 
These Lagrange multipliers are of course 
determined by the normalization conditions, like
\eqn\normal{Z_{0,0}^{+}+Z_{0,0}^{-}+6 Z_{0,1}+12 Z_{0,2}+2 Z_{0,3}=P_{0}.}
As in the pure case, we introduce reduced variables. Here we need 4 ratios $x,y,s$ and $t$ 
defined as,
\eqn\ratiod{
\eqalign{
x&=  u^{-1}  {  y_{0,1} \over (y_{0,0}^{+}y_{0,0}^{-})^{1/2}  } ,  \phantom{AAAA}
y= u^{-1}  {y_{1,0} \over (y_{1,1}^{+}   y_{1,1}^{-})^{1/2} } ,\cr
s&= \big( {y_{0,0}^{+} \over y_{0,0}^{-}} \big)^{1/2}\big( {y_{-} \over y_{+}} \big)^{1/3},  
\phantom{AAAAA}
t= \big( {y_{1,1}^{+} \over y_{1,1}^{-}} \big)^{1/2}\big( {y_{-} \over y_{+}} \big)^{1/3}, 
}}
and for convenience, we also introduce the following averaged weights,
\eqn\aveweight{
\eqalign{
w_{0}&=(Z_{0,0}^{+}Z_{0,0}^{-})^{1/2}, \phantom{AAAA} w_{1}=(Z_{1,1}^{+}Z_{1,1}^{-})^{1/2} ,\cr
w_{2}&=(Z_{2,2}^{+}Z_{2,2}^{-})^{1/2}, \phantom{AAAA} w_{3}=(Z_{3,3}^{+}Z_{3,3}^{-})^{1/2}.}}
The stationarity conditions are then reduced to the following simple form,
\eqn\za{
\eqalign{
Z_{0,0}^{\pm}&=s^{\pm3} w_{0},   \phantom{AAA} 
Z_{0,1}=x w_{0},         \phantom{AAAAA}  Z_{0,2}=x^{2} w_{0}, \phantom{AAA}  
Z_{0,3}=x^{3} w_{0},  \cr
Z_{3,0}&=y^{3} w_{3},     \phantom{AAAA}  Z_{3,1}^{\pm}=y^{2}t^{\pm1} w_{3},\phantom{AA}   
Z_{3,2}^{\pm}=yt^{\pm1} w_{3}, \phantom{AA}  Z_{3,3}^{\pm}=t^{\pm 3} w_{3}, \cr
Z_{1,0}&=y w_{1}, \phantom{AAAAA} Z_{1,1}^{\pm}=s^{2}t^{\pm1} w_{1}, \phantom{AA} 
\tilde{Z}_{1,1}^{\pm}=xys^{\pm1} w_{1}, \cr
 Z_{1,2}^{\pm}&=xs^{\pm1} w_{1}, \phantom{AAA}
\tilde{Z}_{1,2}^{\pm}=x^{3/2}y^{1/2}(st)^{\pm1/2} w_{1}, \cr   
Z_{1,3}^{\pm}&=x^{2}t^{\pm1} w_{1} ,\phantom{AAA}  Z_{2,0}=y^{2} w_{2},   \phantom{AAAA}
Z_{2,1}^{\pm}=ys^{\pm1} w_{2}, \cr
\tilde{Z}_{2,1}^{\pm}&=x^{1/2}y^{3/2}(st)^{\pm 1/2} w_{2}, \phantom{AA}
Z_{2,2}^{\pm}=s^{\pm1}t^{\pm 2} w_{2},\phantom{AA}
\overline{Z}_{2,2}^{\pm}=xyt^{\pm1} w_{2},       \cr
\tilde{Z}_{2,2}^{\pm}&=xyt^{\pm1} w_{2},   \phantom{AAA}
\hat{Z}_{2,2}^{\pm}=x^{1/2}y^{1/2}(st)^{\pm 1/2} w_{2},  \phantom{AA}
Z_{2,3}^{\pm}=xt^{\pm1} w_{2}  .
}}
 Each weight $Z_{i,j}$ is given as the product  of $w_{i}$ by a simple function
 of $x,y,s$ and $t$.
The rules for the $x$ and $y$ variables are simple: The disorder
configuration splits the bonds into those which support a crease and those which
do not. On the bonds with no crease, we assign a factor
$\sqrt{x}$ if the bonds has a fold and 1 otherwise.  On the bonds with a crease, we assign 
a factor $\sqrt{y}$ if the bond has no fold and 1 otherwise. In both cases, the non
trivial factor is assigned if the gauged variable changes sign when crossing the bond.
About the factors of $s$ and $t$, the rules are more subtle. Still 
$s$ and $t$ both measure the symmetry breaking of the frozen parameter.

The weights $w_{0,1,2,3}$ can be expressed as functions of $x,y,s$ and $t$ thanks
to the normalization conditions as follows,
\eqn\za{
\eqalign{
P_{0}&=w_{0}[s^{3}+s^{-3}+6x+12x^{2}+2x^{3}],  \cr
P_{1}&=w_{1}[2y+s^{2}t+s^{-2}t+2xyz+2xyz^{-1}+2xy+2xy^{-1} \cr
&\ +4x^{3/2}y^{1/2}s^{1/2}t^{1/2}+4x^{3/2}y^{1/2}s^{-1/2}t^{-1/2}+x^{2}t+x^{2}t^{-1}],  \cr
P_{2}&=w_{2}[2y^{2}+ys+ys^{-1}+2x^{1/2}y^{3/2}s^{1/2}t^{1/2}+2x^{1/2}y^{3/2}s^{-1/2}t^{-1/2} \cr
&\ +st^{2}+s^{-1}t^{-2}+3xyt+3xyt^{-1}+2x^{1/2}y^{1/2}s^{1/2}t^{1/2} \cr
&\ +2x^{1/2}y^{1/2}s^{-1/2}t^{-1/2}+xt+xt^{-1}]  \cr
P_{3}&=w_{3}[t^{3}+t^{-3}+6y(t+t^{-1})+3y^{2}(t+t^{-1})+2y^{3}].    \cr
}}
 In Eq.\ratiod, the right hand side of each equation is thus a function
 of $x,y,s$ and $t$ only. These equations can be simply solved
 numerically by iteration.

\noindent
%
\fig{Frozen order parameter $F$ versus bending rigidity $K$. 
$F$ changes from $F=0$ to $F\neq 0$. The system shows a 
hysteresis with two separate jumps for two values of $K$ on each
side of $K_{F}$.
The position of $K_{F}$ is determined precisely by comparing
the free energies of both phases, as shown in the next figure.
}{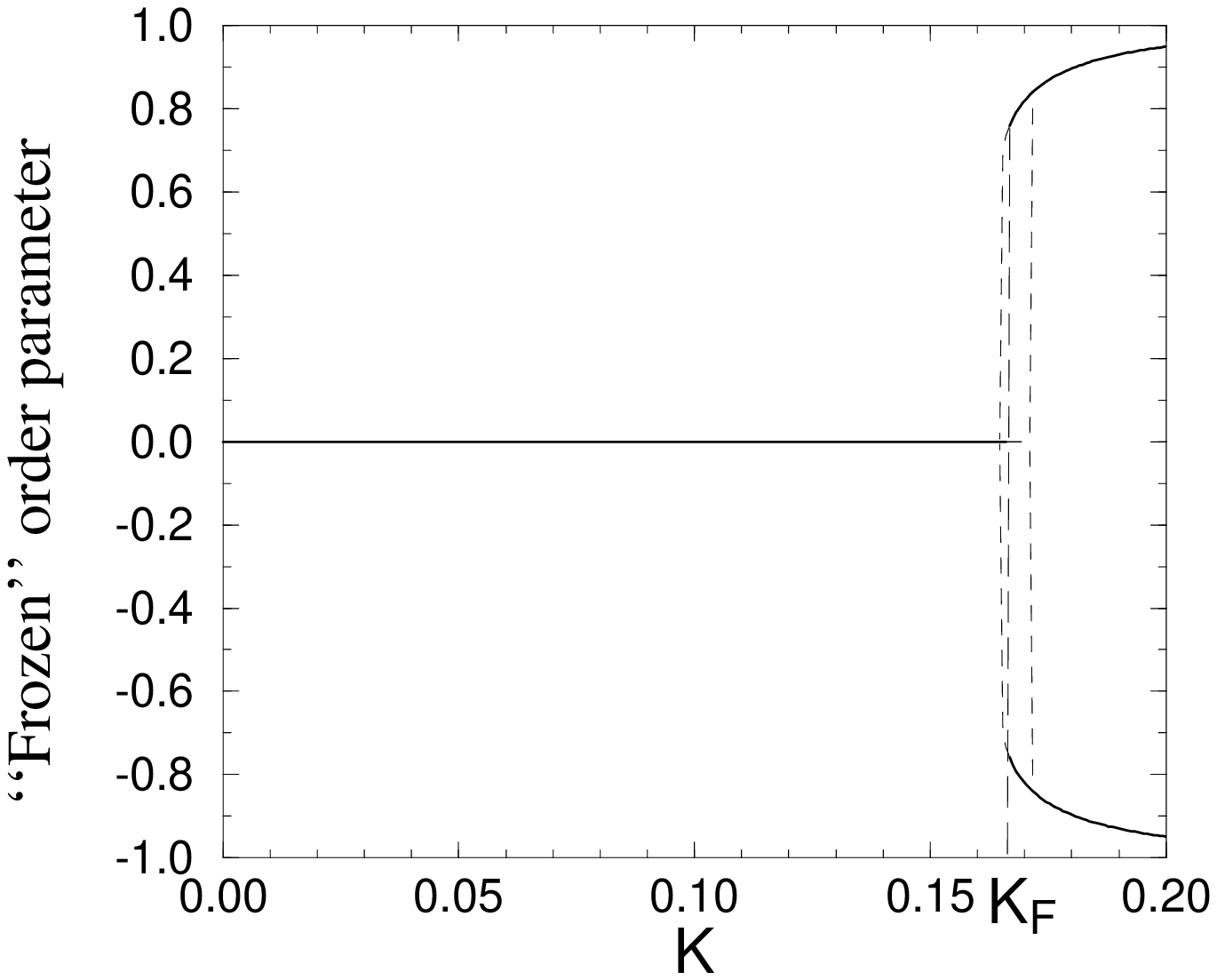}{6.cm}
\figlabel\order
%
\noindent

Hereafter we show the results of this numerical analysis. We first
fix $\alpha =2 $ for the weights $P_{0,1,2,3}$.
We start the iteration for $K=0$ with a fully symmetric 
solution which corresponds to a disordered phase. We proceed to the iteration
until the required precision is reached. We then increase $K$ by $dK$
and restart the iteration. For this next value of $K$, we start the 
iteration from the solution of the iteration for the previous value of $K$.  
This procedure allows to follow the continuous evolution with $K$ of
a given local minimum of the free energy.
We increase $K$ from $0$ to $0.2$ and then decrease it back to $0$.
In this way, if the system has a first order transition with
two local minima of the free energy in competition, the method will 
show a hysteresis. Note that  this iteration procedure is slightly different from 
the natural iteration method that we shall use in section 4.2, 
where we search a solution for the nonlinear 
stationarity equations from different initial assumptions corresponding to 
the different possible symmetries.

\noindent
%
\fig{Free energy $f$ per hexagon versus bending rigidity $K$.
The thin straight line corresponds to a completely frozen phase $F=1$.
This phase is never stable.
The two thick lines correspond to disordered folded phase $F=0$ (for
small $K$) and to the frozen phase $0<|F|<1$ (for larger $K$). 
The two lines cross at the transition point $K_F$. As shown
in the inset, we find $K_F\sim 0.166(1)$.
}{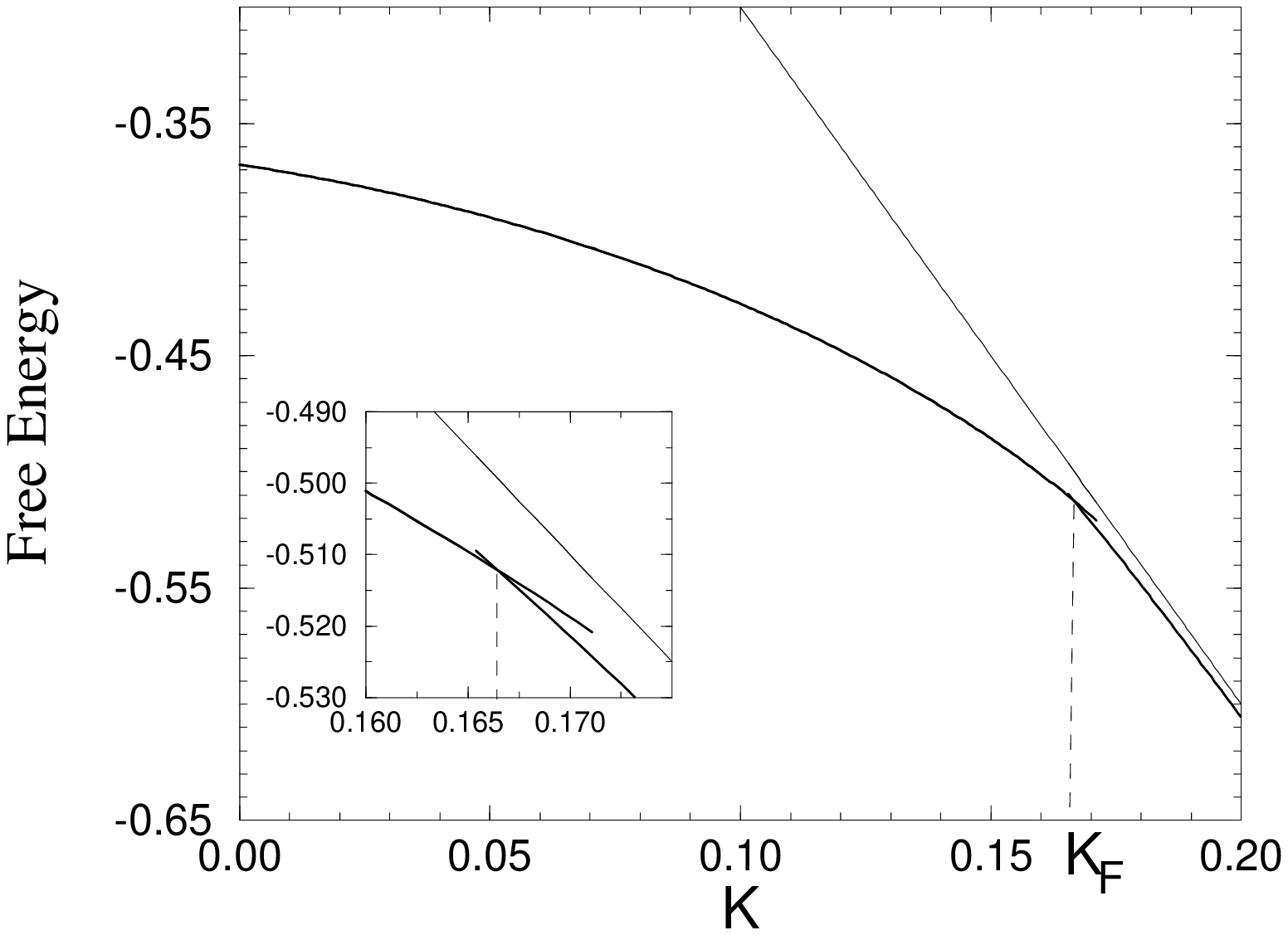}{6.cm}
\figlabel\freeenergy
%
\noindent

In Fig.\order, we show the behavior of the frozen order parameter $F$ as a function of 
the bending rigidity $K$. For small $K (<K_{F})$, $F$ is clearly zero and the system is 
in a disordered folded phase. At $K_F$, the system shows a first order transition 
from this disordered folded phase to a frozen phase $F\neq 0$. The value of $\vert F \vert $
is strictly less than 1, thus the system is only partially frozen.
When the iteration is performed with first increasing $K$ and then decreasing
it back to zero, we see a clear hysteresis with two jumps
on both sides of $K_F$.  
The value of the transition point $K_{F}$ can be fixed precisely by comparing the 
value of the free energies for both phases.
This is shown in Fig.\freeenergy.
The hysteresis allows us to see clearly the crossing of the two free
energy lines corresponding to both phases $F=0$ and $F\ne 0$.
Indeed the system stays for some time after the transition point in the wrong
metastable state. As shown in the inset of the figure, the transition occurs
at $K_F\sim 0.166(1)$.

\noindent
%
\fig{Gauged two point function $\overline{<\eta_{1}\eta_{2}>}
= - {\rm internal}$ ${\rm energy}/K$ versus $K$. We see here also a clear evidence of 
first order transition with a hysteresis.
}{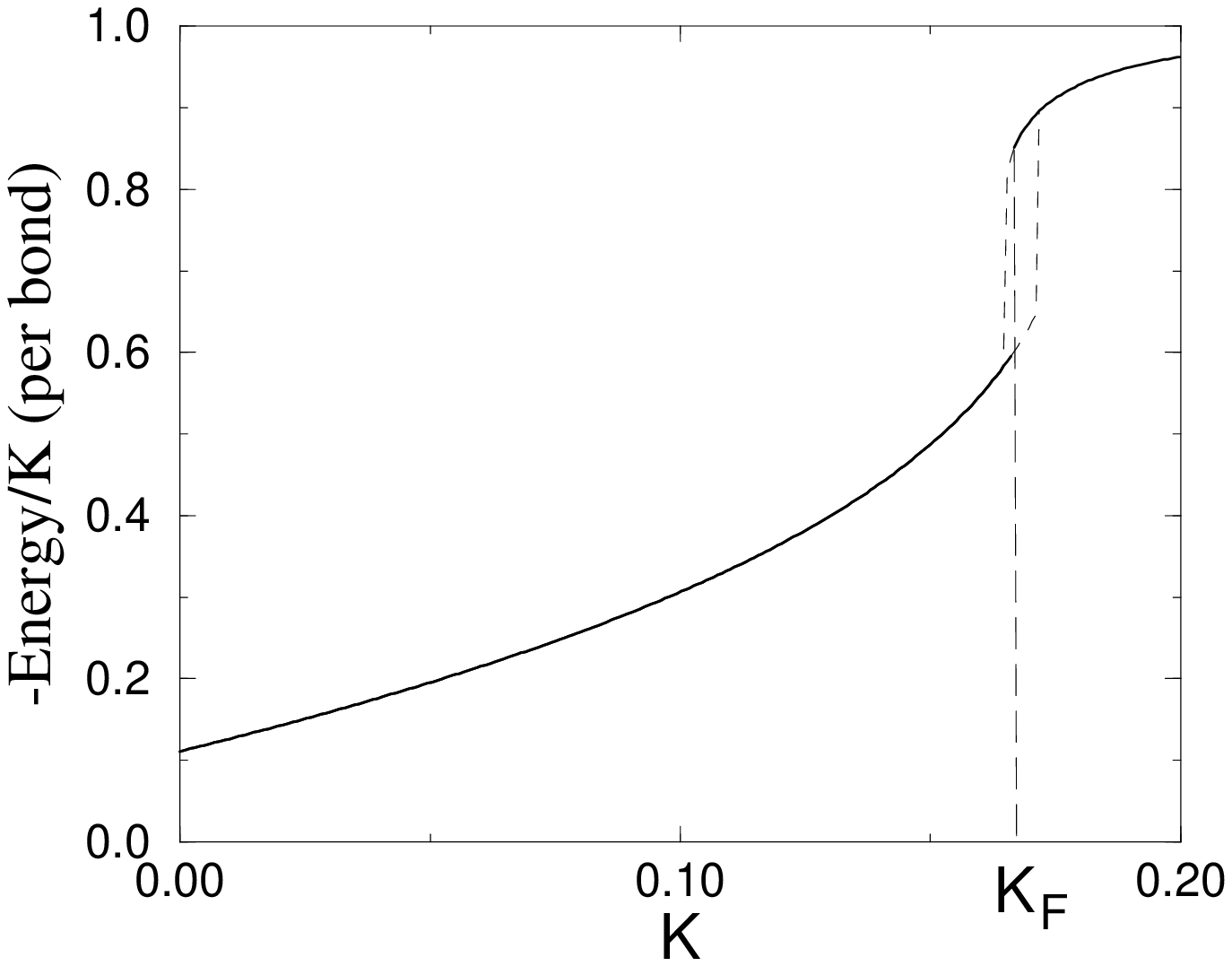}{6.cm}
\figlabel\energy
%
\noindent

Fig.\energy\ shows the behavior of the two-point function 
$\overline{<\eta_{1}\eta_{2}>}=- {\rm internal}$ ${\rm energy}/K $
 versus $K$. It also shows a clear evidence of
first order transition with a hysteresis in the results of the
iteration procedure. At $K=0$, the disorder variables $\tau_{i}$
 and spin variables $\sigma_{i}$ decouple. The value of the function is then
$\overline{<\eta_{1}\eta_{2}>}=\overline{\tau_{1}\tau_{2}}<\sigma_{1}\sigma_{2}>_{{\rm at} K=0}
=(-1/3)\times(-1/3)=1/9$, as found here (the value $-1/3$ is easily obtained
from the analysis of the pure case at $K=0$ of section 3.2 \CGP).

\noindent
%
\fig{Frozen order parameter $F$ versus $K$ for several values of $\alpha$
(we only show here the case $F\ge 0$).
The jump in the order parameter becomes smaller as $\alpha$ becomes larger.
The continuous character of the transition is recovered at $\alpha=\infty$ (Pure 
antiferromagnetic system. For intermediate $\alpha$, the iteration gives rise to a
hysteresis.)
}{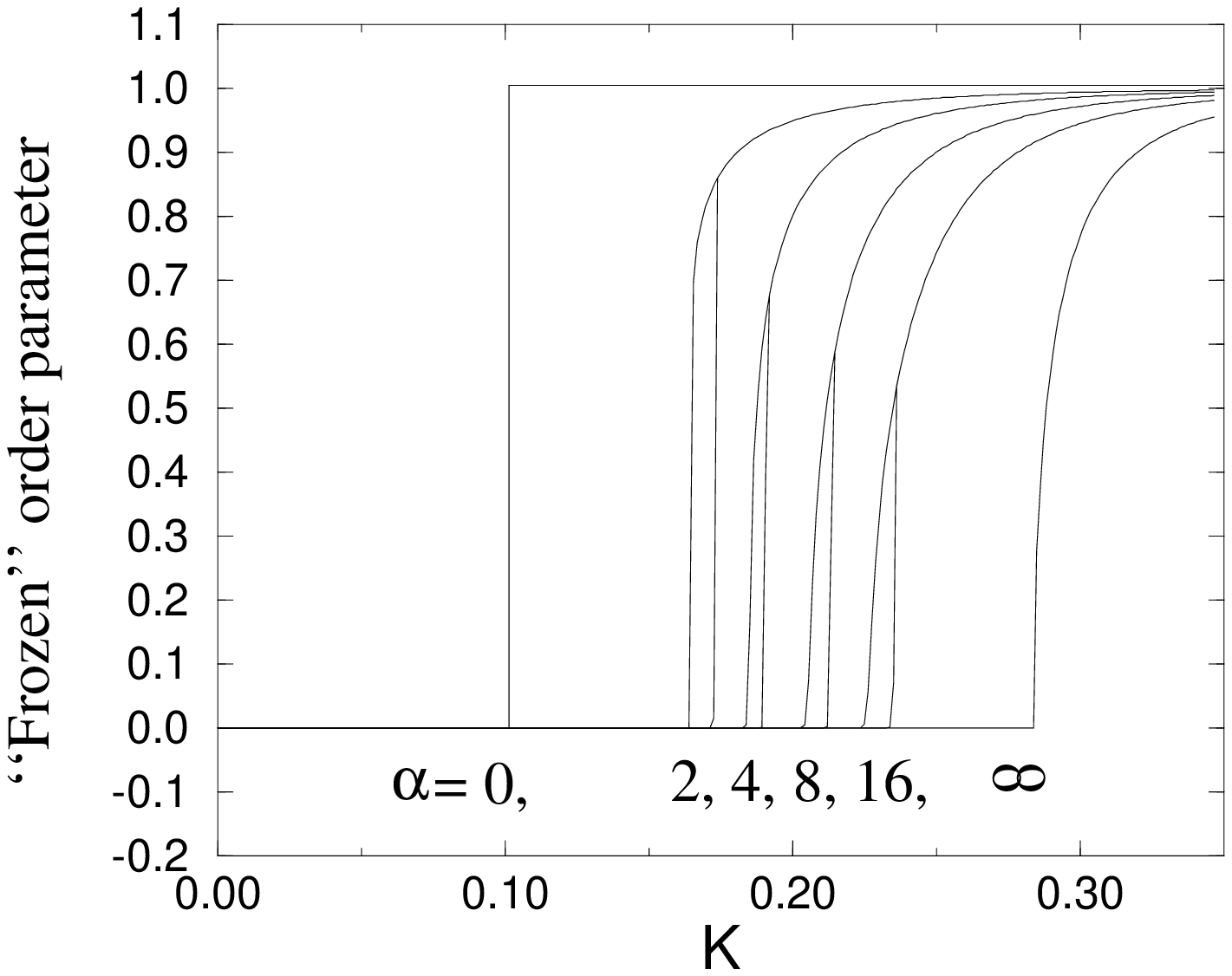}{6.cm}
\figlabel\valpha
%
\noindent

 Next, we studied the system  for several
 values of the parameter $\alpha$ for the disorder weights. 
As has been explained previously, $\alpha=0$ means that there is no crease in the 
system and $\alpha=\infty$ corresponds to the pure antiferromagnetic system.
{}From the previous analysis, we know that the system shows a first order transition 
for both $\alpha=0$ and $\alpha=2$ above. At $\alpha =0$ (where $F$=$M$) the
transition is from $F=M=0$ to $F=M=\pm 1$ \DIGS. At $\alpha =2 $, the discontinuity
is smaller with $\vert F\vert <1$ in the frozen phase. We also know
that the transition becomes continuous and second order at $\alpha=\infty$ \CGP.
The transition point is at $K=0.284$ as obtained before.
{}From the results in Fig.\valpha, we see that the discontinuity of the transition becomes 
smaller as we increase the parameter $\alpha$. The continuity of the transition 
seems to  be recovered only at $\alpha=\infty$, although it is difficult to determine 
whether the transition is of first order or of second order when the discontinuity 
becomes too small.

\subsec{(K,h)-Phase diagram}

\noindent
%
\fig{Phase diagram in the $(K,h)$ plane for the fully 
disordered system (Eq.\hammattis). 
First order lines separate the four phases: 
(1) Disordered folded phase with $M\sim 0$, $F=0$, 
(2) Completely Flat Phase with $M=1$ and $F=0$, 
(3) Flat Phase with $0<|M|<1$ and $F=0$ and
(4) Frozen phase with $M=0$ and $|F|>0$.}{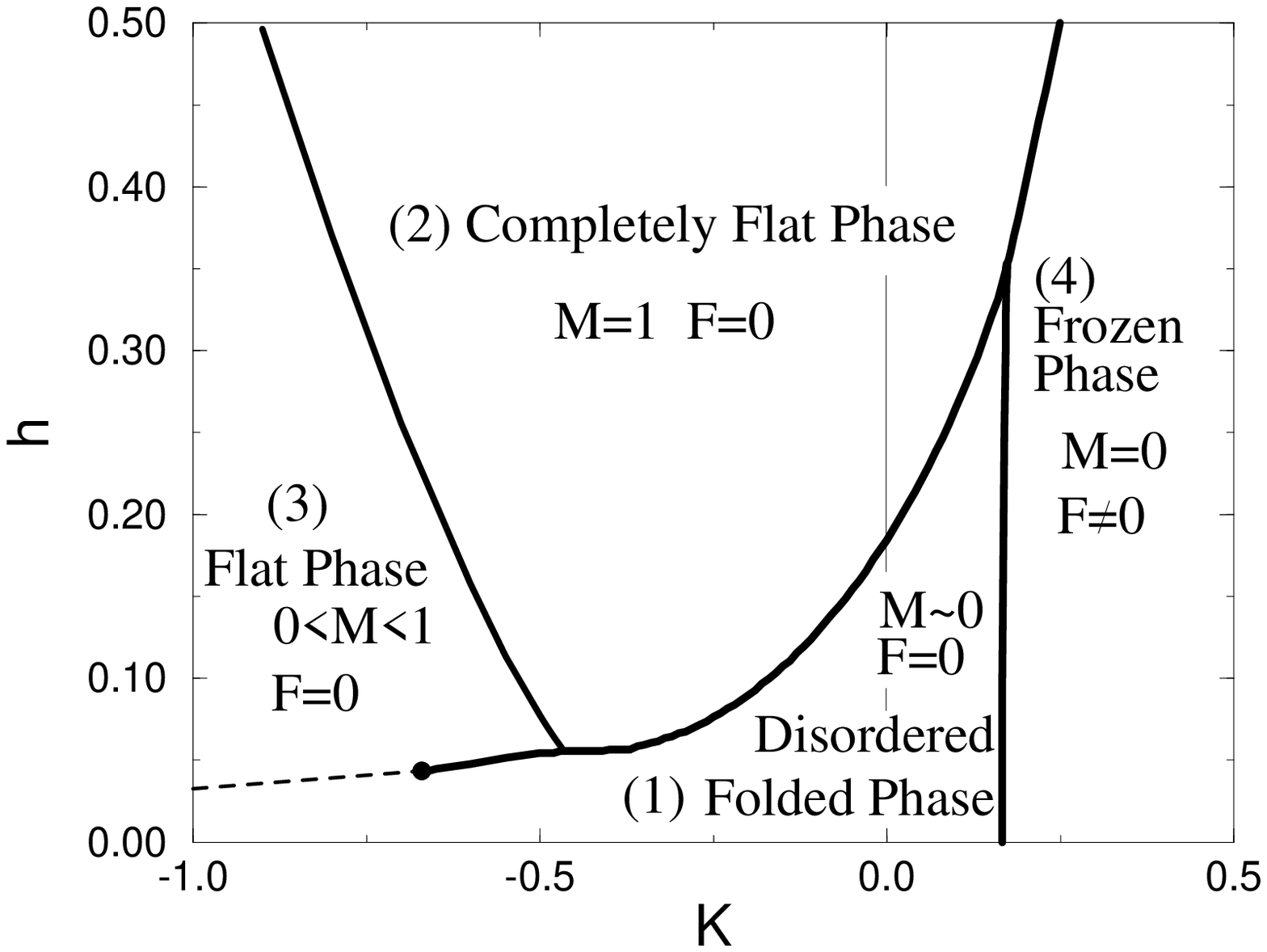}{7.cm}
\figlabel\pdomodel
%
\noindent

Let us now turn to the analysis of the whole phase diagram in the $(K,h)$-plane.
It is of course symmetric with respect to the $h=0$ axis. It 
is shown in Fig.\pdomodel\ for $h\geq 0$ as obtained from the
CVM stationarity equations solved by the natural iteration method \K.
Here we have set $\alpha=2$ again and the results were obtained with a set
of $22\times 22$ independent weights, i.e. without making
any assumption on the symmetries of the different phases, except in 
the initial conditions of the iteration.
At sufficiently large values of $h$ and for $h>2K$, the completely flat phase with $M=1$
 is stable with respect to both the disorder and the thermal fluctuations. 
At sufficiently large values of $K$ and for $h<2K$, the system is in a frozen phase 
with $M=0, M_{\rm st}=0$ and $F\sim 1$. 
There is a first order transition line between these two phases, 
which is roughly given by $h=2K$. The position of the line can be obtained by 
requiring that the free 
energy of the completely flat phase and that of the frozen phase take the 
same value, i.e. by solving 
the equation $f_{\rm c.flat}=f_{\rm frozen}$. 
About $f_{\rm c.flat}$, the bending 
energy contribution per hexagon is estimated as $-3K\overline{<\tau_{1}\tau_{2}\sigma_{1}\sigma_{2}>}
=-3K\overline{(\tau_{1}\tau_{2})}=-3K\times(-1/3)=K$, while the 
entropy vanishes in the absence of local excitations.  We thus get the exact free energy
$f_{\rm c.flat}=-2h+K$.
The estimation of $f_{\rm frozen}$ is more difficult and we simply assume that $f_{\rm frozen}\simeq -3K$
as in a completely frozen phase, 
because the frozen order parameter $F$ is almost saturated to $1$. {}From these estimations, 
we obtain the transition line $h\simeq 2K$, which is what we indeed observe.

For smaller $K$ and $h$, the system is in a disordered folded phase with $F=0$ and $M\sim 0$.
As for the pure system, $M$ does not vanish exactly for $h>0$ but still remains very small.
This might be an artifact of the CVM approximation \CGP.
At $K=0$, the spin variables $\sigma$ are decoupled from the disorder variables $\tau$
and the fully disordered system is the same 
as the pure system. It has a first order transition point at $h\sim 0.184$ \DIGS.
For $h=0$, the above results confirm those of previous section with a transition
at $K=K_F$.

\noindent
%
\fig{Lowest energy spin configurations $\{\sigma_{i}\}$ for each type of disorder 
configuration. Folds are indicated by thick lines and creases by dashed lines.  
On the left hand side, we show the disorder configurations, 
in the center the corresponding lowest energy state, which violates the 
local folding constraint for the disorders with 1 and 2 creases, and
on the right hand side the lowest energy states which preserve the constraint,
together with their degeneracy. 
}{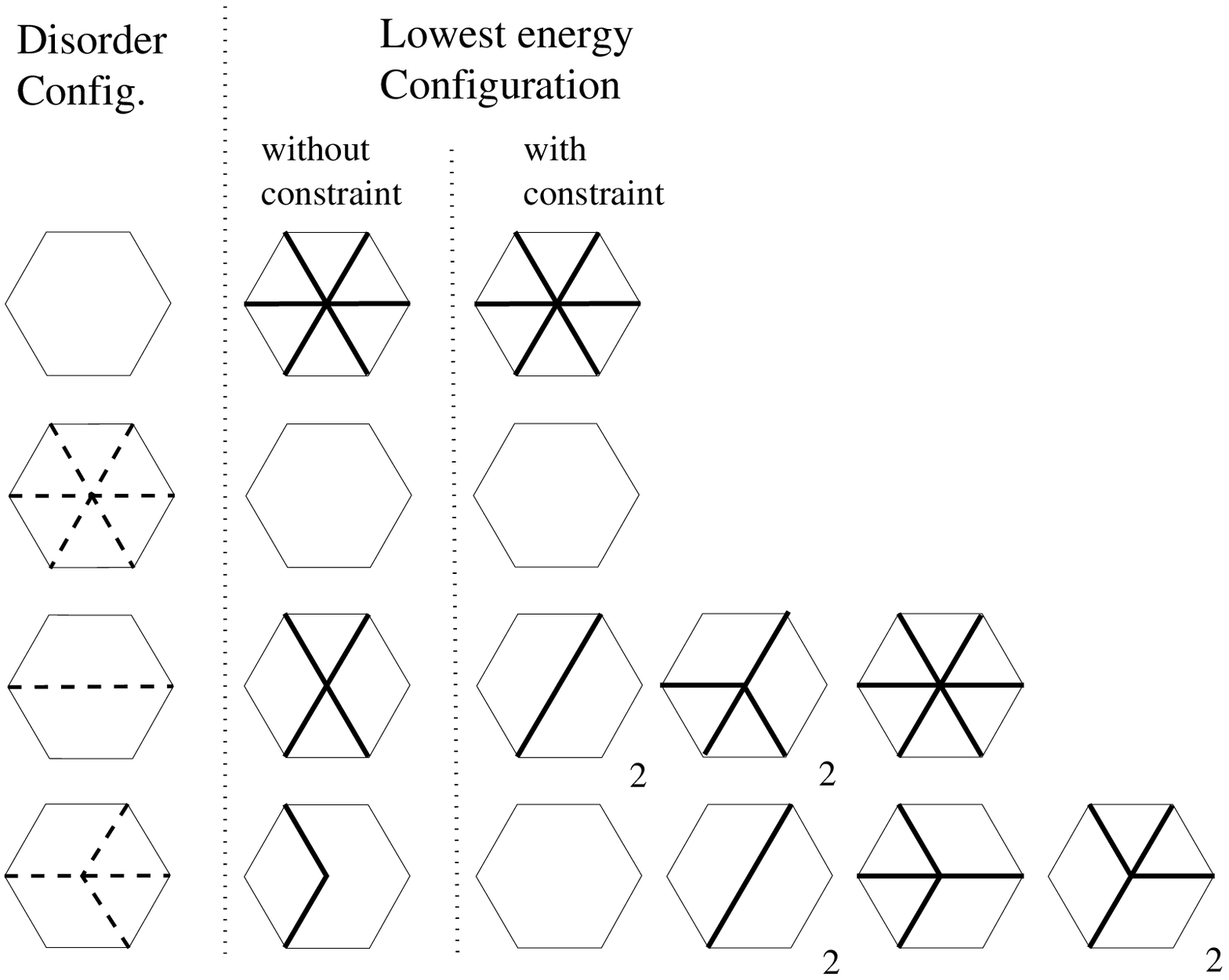}{6.cm}
\figlabel\frust
%
\noindent

We have also studied the case of negative $K$, although it is not
very physical. Still, it presents interesting features in view
of our further study of other types of disorder in Section 5.
In the usual Mattis Model without constraint and for $h=0$, the spins develop at
large enough negative $K$ an ``antiferromagnetic''-like order in the gauged
variable $\eta_i$ with a ground state $\eta_i=(-1)^{i-1}$ in the limit $K=-\infty$.
Here such order cannot be reached in general due to the constraint on $\sigma$,
hence on $\eta$.
If the disorder has 0 or 3 creases, then the ground state $\eta_i=(-1)^{i-1}$
can be reached and is the unique ground state spin configuration.
On the other hand, if it has 1 or 2 creases, it cannot be reached 
and we are led to several lowest energy spin configurations, as shown 
in Fig.\frust . 
At $K=-\infty$ and $h=0$, the actual ground state will thus be degenerate with
frustrations in the system which might prevent the emergence of a true
``frozen antiferromagnetic'' order. The system is thus always disordered in contrast
with the pure case where an antiferromagnetic order had developped.

\noindent
%
\fig{Gauged two point function $\overline{<\eta_{1}\eta_{2}>}=
\overline{<\sigma_{1}\tau_{1}\sigma_{2}\tau_{2}>}$ 
along the  $K$-axis $(K<0, h=0)$.}{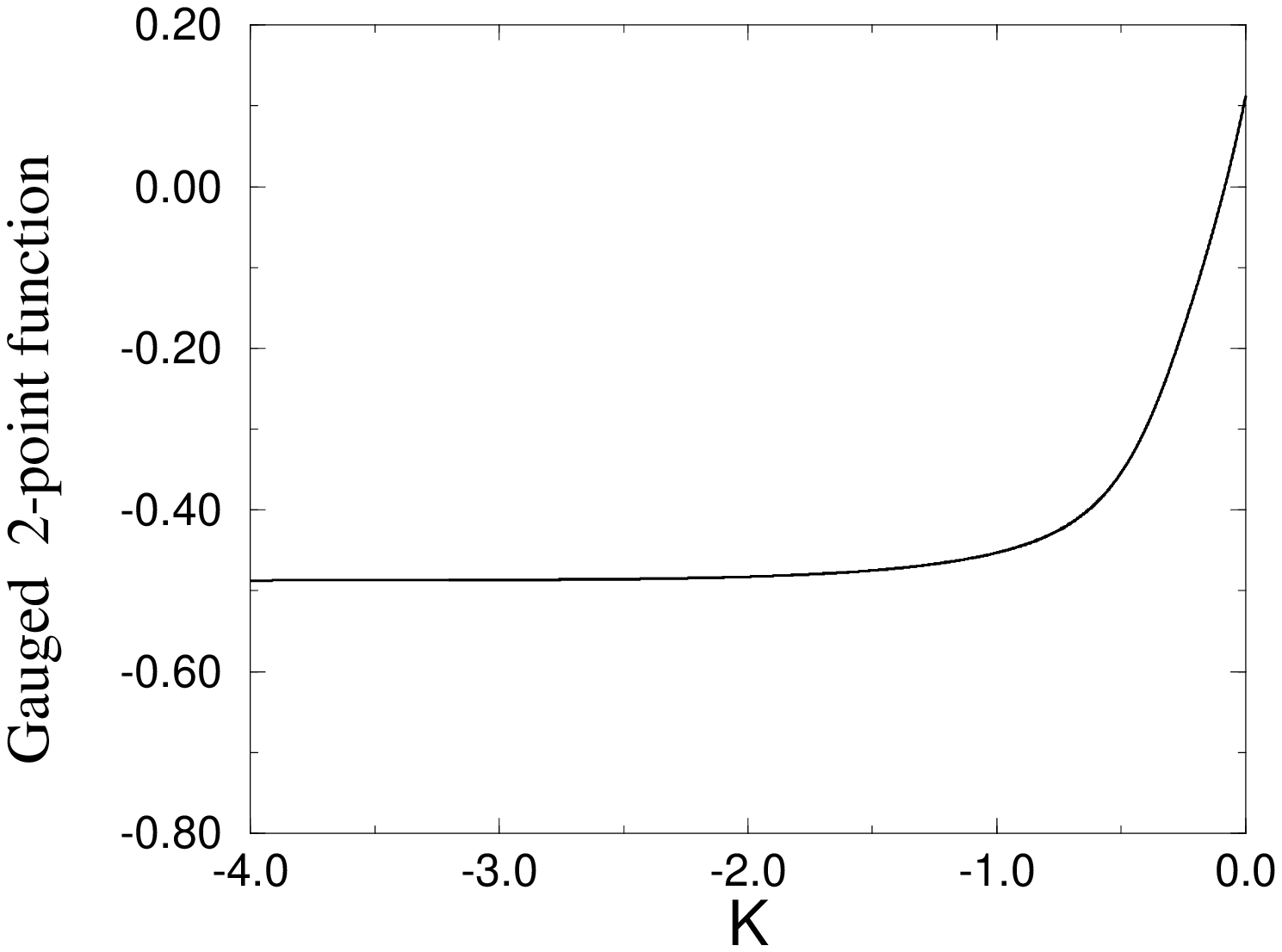}{5.0cm}
\figlabel\cfun
%
\noindent

{}From the analysis of the lower energy states, we can easily compute
the two-point correlation at $K\rightarrow -\infty$,
$c=\overline{<\sigma_{1}\tau_{1}\sigma_{2}\tau_{2}>}$ 
for the gauged variable (i.e. the  internal energy $/K$). 
For a disorder with 0 or 3 creases, we have
$\sum_{i\ {\rm around}\ v} <\sigma_{i}\tau_{i}\sigma_{i+1}\tau_{i+1}>=-6$ in the 
lowest energy state. For a disorder with 1 or 2 creases,  
we have $\sum_{i \ {\rm around} \ v} <\sigma_{i}\tau_{i}\sigma_{i+1}\tau_{i+1}>=-2$ 
for all the lowest energy states.
The averaged value is thus estimated as 
$c=\overline{<\sigma_{1}\tau_{1}\sigma_{2}\tau_{2}>}=(-6\, (P_{0}+P_{3})-2\, (P_{1}+P_{2}))/6 =-19/39\sim -0.48$.
This is what we observe (see fig \cfun .). The two-point function does
not show any discontinuity in all the negative $K$ regime, which confirms 
the absence of transition in this regime (at $h=0$).

Finally, we included in Fig. \pdomodel\ the results of the CVM analysis
for negative $K$ and arbitrary $h$. We can see the emergence of a new partially flat
phase with $0<\vert M\vert <1$. The nature of the transition
between this phase and the disordered folded phase is unclear, in
particular because in the latter phase, $M$ is not exactly zero within
the CVM approximation. We then see a limiting point 
(black circle in Fig. \pdomodel ) below which the magnetization
does not present any longer a discontinuity between the two phases.
This also might be an artifact of the CVM, in which case the true
transition line should be continued to lower values of $K$ (dashed line).

\newsec{Other Models}

In this section, we complete our study by considering other variants of the disorder. 
As discussed in section.2, there are several possibilities for the 
choice of $K_{ij}$. In order to appreciate the importance of the local folding 
constraint \locons\ on the disorder face variables $\tau_{i}$, we 
will study the model Hamiltonian \hammattis\ without the local folding 
constraint on $\tau_{i}$.
Next we will study the Edwards-Anderson model with the local folding constraint on 
the spin variables $\sigma_{i}$ and a bending term $K_{ij}$ given by 
$K_{ij}=K\tau_{ij}$ with a random variable $\tau_{ij}=\pm 1$ 
on each bond (ij).  
We refer to the former case as the model (2) and to the latter case as the model (3).
The difference between models (2) and (3) is simply the possibility in the model (3)
of having vertices with an odd number of surrounding creases.
We also refer to the original model \hammattis\ with the folding constraint on $\tau_{i}$ 
as the model (1). 
\noindent
%
\fig{One of the lowest energy spin configurations $\{\sigma_{i}\}$ 
for each type of disorder configurations. Folds are shown by thick 
lines and creases are by dashed lines.  
The degeneracies for the disorder configurations are indicated.
We also give the value of the corresponding minimal energy.
}{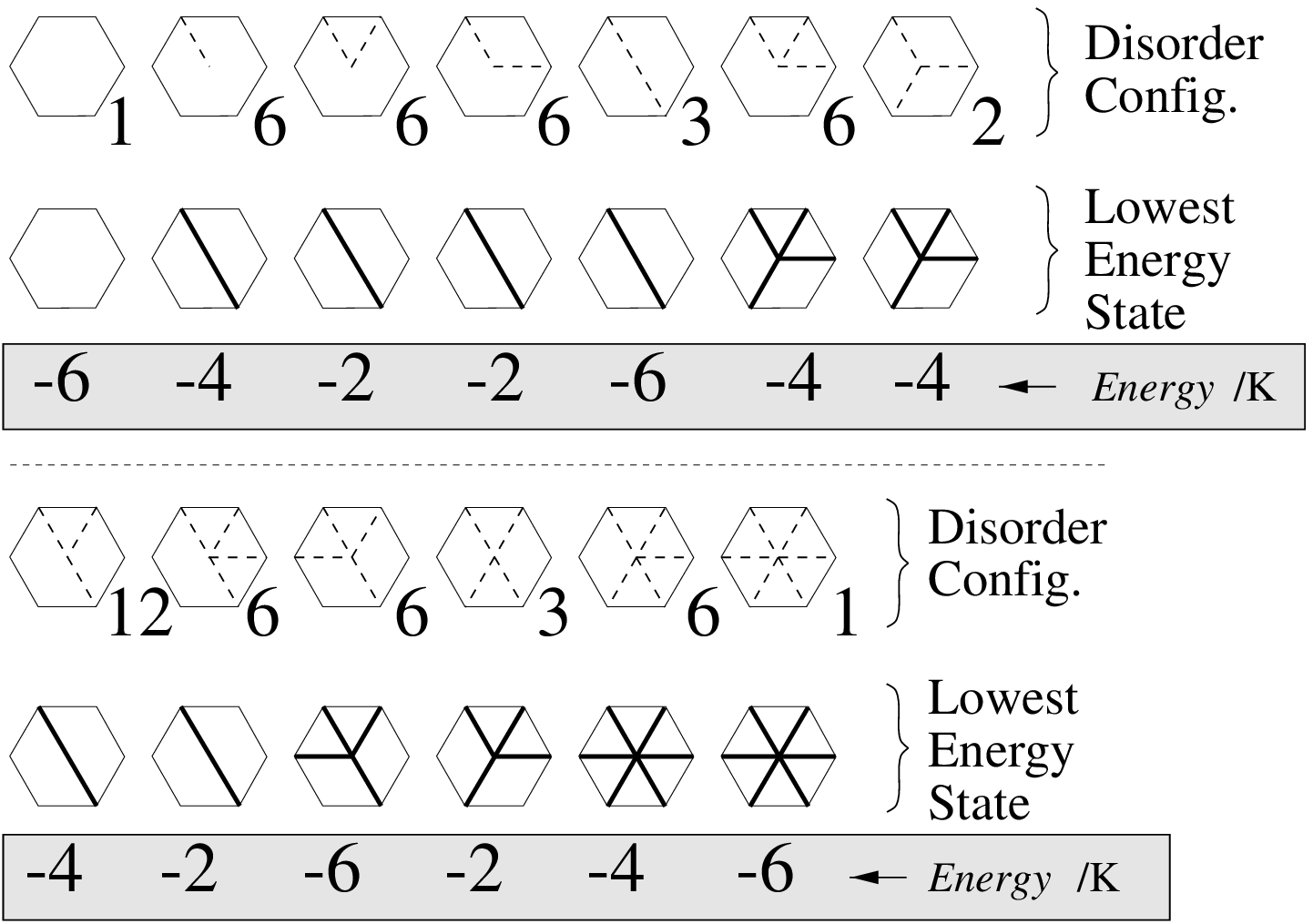}{8.cm}
\figlabel\frusm
%
\noindent
Before we discuss the corresponding $(K,h)$ phase diagrams, let us discuss the two-point 
function of the system at $h=0$ and large $K$. Clearly in this limit, for
a fixed disorder configuration, the $\sigma_i$ variables will tend to minimize
the energy $-K_{ij}\sigma_i\sigma_j$, i.e. will tend to maximize the
overlap with the disorder configuration in terms of folded bonds. In other
words, the system wants to create a fold ($\sigma_i\sigma_j=-1$) whenever
a crease exists ($K_{ij}/K=-1$), and no fold otherwise. 
For an arbitrary environment of $K_{ij}$ around a vertex, we can easily
find one corresponding lowest energy state for the $\sigma$ variables around
the vertex. There are in general several such states. In Fig. \frusm ,
we have displayed all disorder environments together with one of
the corresponding lowest energy state.
It is interesting to notice that the disorder configurations can be arranged
in three categories:
\item{(i)} Those with an even number of creases and which satisfy the folding
constraint. The lowest energy state is unique and has energy $-6K$.
\item{(ii)} Those with an even number of creases but which do not satisfy the folding
constraint. The corresponding minimal energy is $-2K$ in this case.
\item{(iii)} Those with an odd number of creases. The corresponding minimal 
energy is $-4K$ in this case.

Of course, in the model (2), only vertices of type (i) and (ii) are allowed while
in the model (3), all vertices can appear.

At $K\rightarrow \infty$ and $h=0$, we can thus estimate the two-point function 
$\overline{<\eta_{1}\eta_{2}>}= \overline{<\sigma_{1}\tau_{1}\sigma_{2}\tau_{2}>}$
for the model (2) and $\overline{<\sigma_{1}\tau_{12}\sigma_{2}>}$ for the model (3)
by averaging over all disorder environments the corresponding
minimal energy.
All (allowed) disorder environments are now equiprobable. Taking into account
the appropriate degeneracies under rotations, we get for the model (3)
\eqn\internal{\eqalign{
&\overline{<\sigma_{1}\tau_{12}\sigma_{2}>}=(-6\times 1 -6\times 1 -4 \times 6 -2 \times 6 
-2 \times 6 -6 \times 3 -4 \times 6 \cr 
&-4 \times 2  -4 \times 12 -2 \times 3 -2 \times 6 
-6 \times 6 -4 \times 6)/(6\times 64) =59/96 \sim 0.614. }}
For the model (2) we find easily:
\eqn\internals{\overline{<\sigma_{1}\tau_{1}\sigma_{2}\tau_{2}>}=18/32\sim 0.562.}

\noindent
%
\fig{Gauged two-point function $\overline{<\sigma_{1}\tau_{1}\sigma_{2}\tau_{2}>}$
for the model (2) (solid line) and $\overline{<\sigma_{1}\tau_{12}\sigma_{2}>}$ 
for the model (3) (dashed line) versus $K$ for $h=0$.}{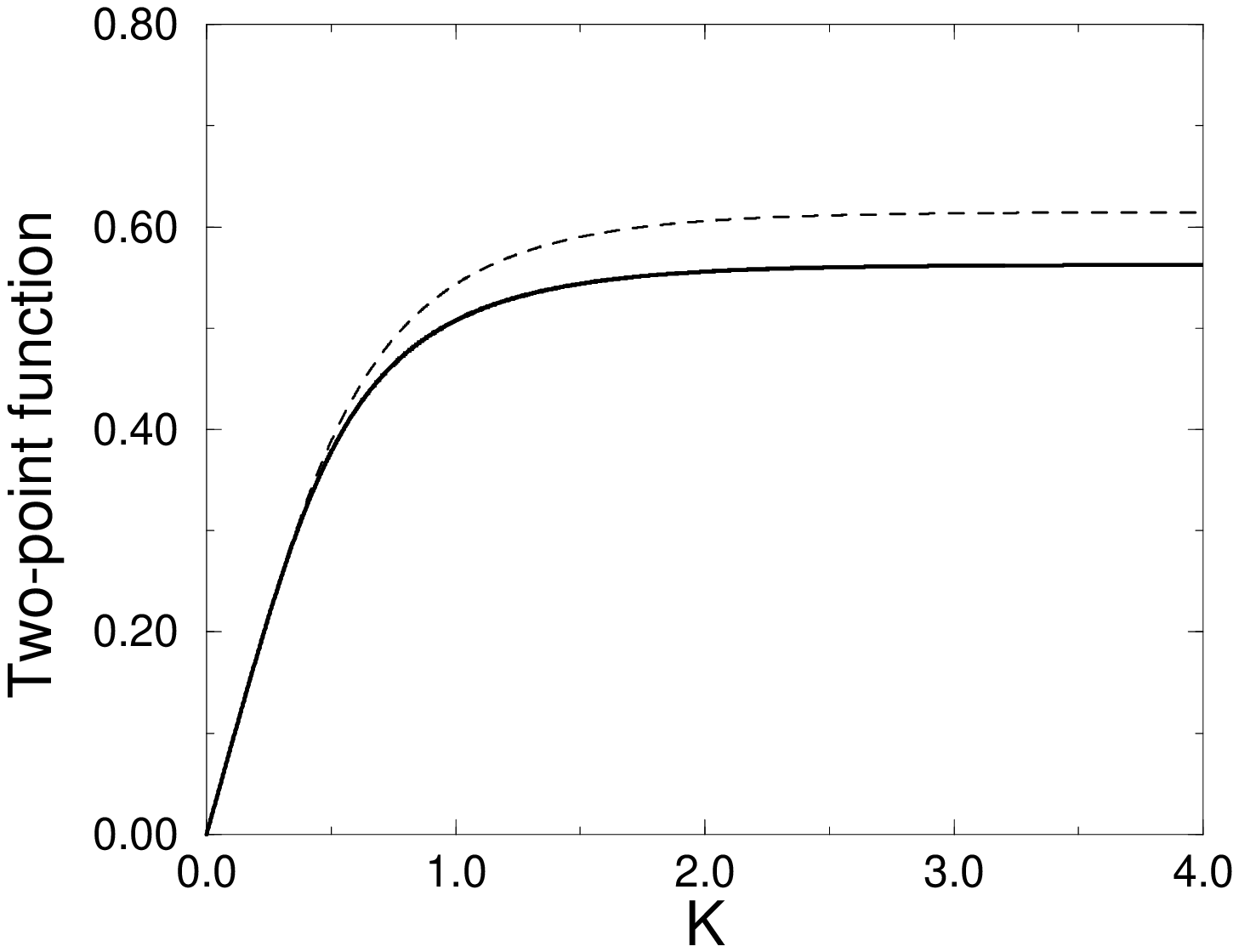}{5.0cm}
\figlabel\cfunst
%
\noindent
Note that the above calculation assumes that a lowest energy state
can be constructed globally out of these local lowest energy configurations.
This assumption is acceptable within the CVM approximation at least.
We show on Fig. \cfunst\ the two-point function for the models (2) and (3)
as obtained from the CVM They do not display any discontinuity
and tend to the values calculated above at large $K$.

Note also that in some sense, the model (3) is less frustrated than 
the model (2) since a better overlap with a constrained $\sigma$ configuration 
can be obtained for those frustrated disorder environments with an odd
number of creases.

\noindent
%
\fig{Phase diagram in the $(K,h)$ plane for the model (2).
We find three phases: (1) a disordered 
phase with $F=q=0$, (2) a completely flat phase with $M=1$ 
and $F=q=0$, and (3) a frozen flat phase with $0<M<1$, $q>0$ and $F>0$.
The solid line represents a first order transition line and the dashed
line a continuous transition line.
}{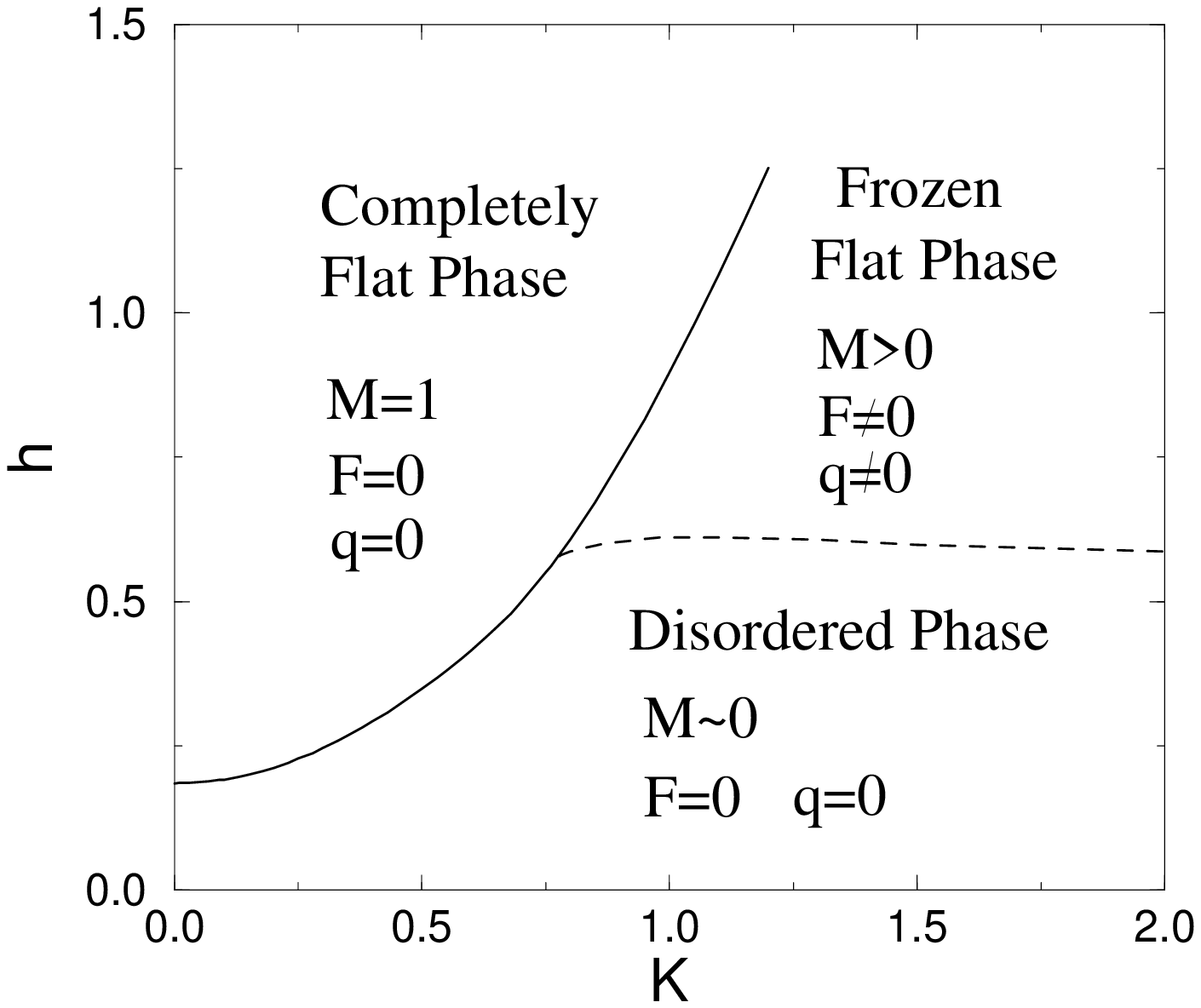}{6.cm}
\figlabel\pdsmodel
%
\noindent

\nobreak
%
\fig{Phase diagram in the $(K,h)$ plane for the model (3).
Two phases are separated by a first order transition line: (1) a frozen
flat phase with $0<M<1$ and $q>0$) and 
(2) a completely flat phase with $M=1$ and $q=0$.
The disordered folded phase with $M=q=0$ is recovered only at $h=0$.
}{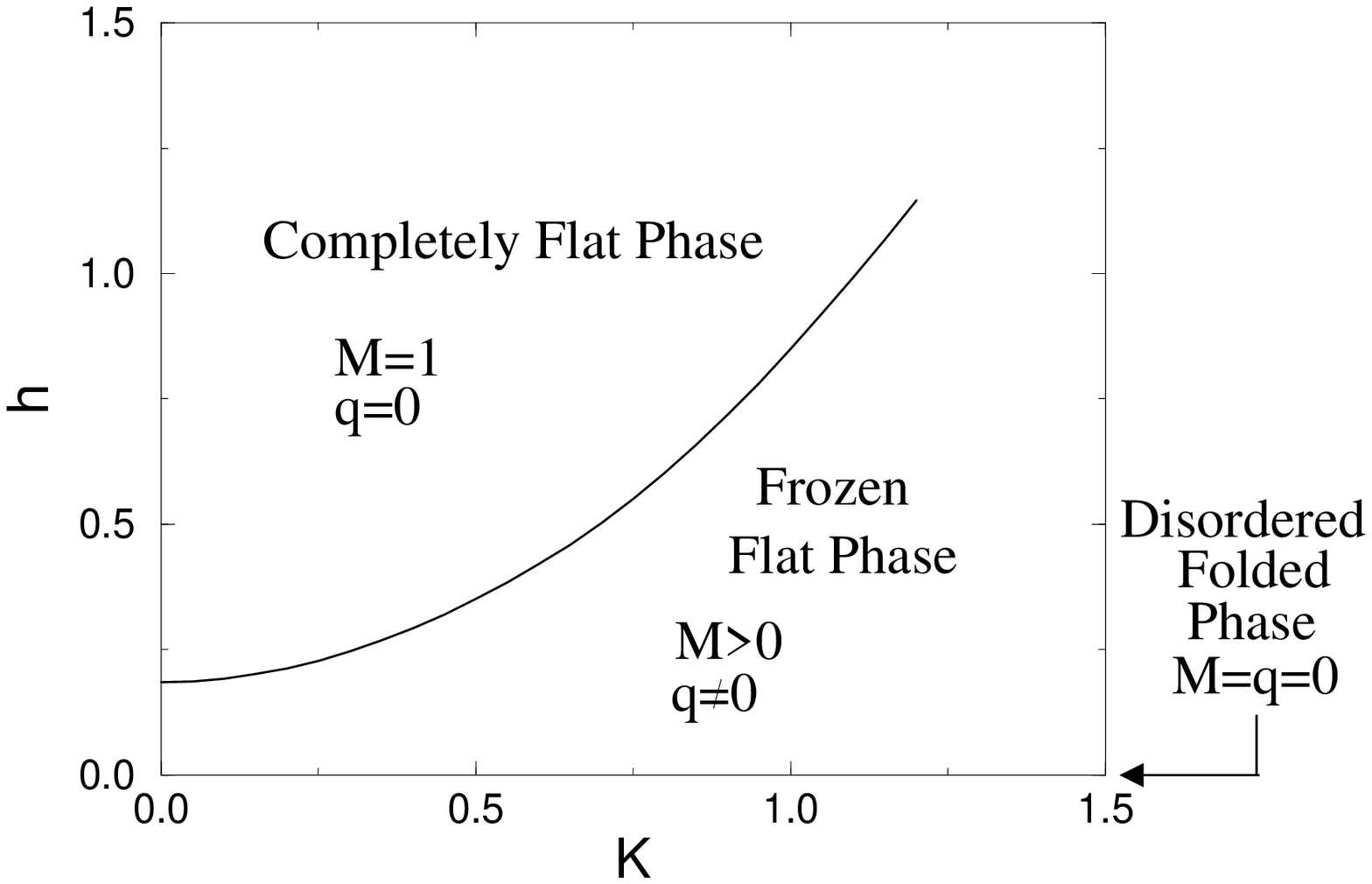}{6.cm}
\figlabel\pdtmodel
%
\nobreak

We have studied the phase diagrams of both models (2) and (3) within
the CVM approximation by use of the natural iteration method \K.
In order to characterize each phase, in addition to the previous order parameters 
$M$ and $F$ we will also use the spin-glass order parameter $q$ defined
in Eq.\spgl , generalized to the case where $M\neq 0$: 
\eqn\mspgl{q={1 \over N_t}\sum_{i} [\overline{\langle \sigma_i \rangle^{2}}-
\overline{\langle \sigma_i \rangle}^{2}] .}
In the model (2), both $F$ and $q$ measure the frozen character of the phase.
They are expected to be zero (resp. non-zero) simultaneously for our
choice of disorder distribution.
In the model (3) with no face $\tau_i$ variables, $F$ is not defined any longer
and we will use $q$ as a measure of the frozen character of the phase.
The phase diagram for the model (2) is shown in Fig.\pdsmodel\ and that for 
the model (3) in Fig.\pdtmodel.
They are of course symmetric with respect to the $K=0$ axis but also 
with respect to the $h=0$ axis. This is because the transformation
$\tau_i\to (-1)^{i-1}\tau_i$ in the model (2) and $\tau_{ij}\to -\tau_{ij}$
in the model (3) exchange equally probable disorder environments and
simply change $K$ into $-K$.  
We only display the phase diagrams for $K>0$ and $h>0$.

At $K=0$, both systems are identical to the pure system and undergo a first order 
transition at $h\sim 0.184$ to a completely flat phase. 
This completely flat phase with $\vert M\vert =1$ and $q=0$ is stable for all
$K$ at large enough $h$ above a line which is almost the same
for the two models up to the tricritical point of Fig.\pdsmodel .
At $h=0$, both system remain in a disordered folded phase with
$M=q=0$ for any value of $K$. This is different from the model (1)
where we had a transition to a frozen phase at $K=K_F$. 
The absence of a frozen phase at $h=0$ is again due to the presence
of frustration in the system leading to several competing lowest energy states.

For fixed (large enough) $K$ and increasing $h$, the models (2) and (3)
display different behaviors. As far as $q$ is concerned, the model (3)
develops a non-zero value of $q$ for any $h>0$. On the other hand,
$q$ remains zero in the model (2) until a critical value of $h$
is reached where a continuous transition to $q\neq 0$ occurs.
For both models, the ``frozen'' phase with $q\neq 0$ also has $0<\vert M\vert < 1$,
and is thus partially flat. As to the $q=0$ phase of the model (2), we see
a non-zero value of $M$ which is indeed non negligible close to
the continuous transition line. Still we cannot exclude that this could be
again an artifact of the CVM approximation. Indeed, by continuity from 
the $K=0$ line, we would rather expect $M=0$ everywhere in this phase.
This issue is thus not fully solved.
Finally, the absence of the $q=0$ phase in the model (3) (except for $h=0$)
might also be interpreted again as an indication of a weaker frustration
as compared to the model (2).

\newsec{Discussion and Concluding Remarks}

 In this paper we have studied the folding of the triangular 
lattice in the presence of a quenched random bending rigidity $K_{ij}=\pm K$
and a magnetic field $h$. We have considered  three types of quenched randomness
(1) $K_{ij}=K\tau_{i}\tau_{j}$ with face random variables 
$\tau_{i}\pm 1$ subject to the folding constraint \locons; (2)  
$K_{ij}=K\tau_i\tau_j$ without the folding constraint on the $\tau_{i}$'s; (3) $K_{ij}=K\tau_{ij}$ with 
a bond random variable $\tau_{ij}=\pm 1$. In case (1), the folding constraint on 
the disorder variables was introduced to describe a particular type of ``physical'' disorder
supposed to mimic that induced in a randomly crumpled surface, here 
in the context of the folding of the triangular lattice. 
Applying the cluster variation method generalized to random systems, 
we have studied the phase diagrams of the three models (1),(2) and (3) and 
their phase transitions.  
The phase diagrams for each case are depicted in figs.
\pdomodel, \pdsmodel\ and \pdtmodel\ respectively. The most important 
difference between the model (1) and the models (2) and (3) is that, in
the absence of magnetic field, a frozen phase is found only in the model (1),
for large enough $K$. In this phase, the configuration of the triangular 
lattice is trapped in the randomly oriented phase characterized by the 
configuration of the disorder variables $\{\tau_{i}\}$.
The models (2) and (3) do not present such frozen order at $h=0$. Indeed,
these models, where the quenched randomness is not 
constrained, have strong frustrations
because the constrained spins describing the normals to folded configuration 
fail to be in the ``virtual'' lowest energy ground state dictated by 
the unconstrained disorder, even if the coupling constant $K$ becomes large. 
For $h>0$, a frozen phase is recovered in the models (2) and (3). We find several 
first order or continuous transition lines between 
the frozen phase and a completely flat phase or a disordered folded phase.

At last we make one comment about previous studies on another spin model describing a
polymerized membrane with quenched random spontaneous curvature [\xref\BMPS,\xref\ACB],
with Hamiltonian
\eqn\hamprevious{
{\cal H}=-\sum_{ij}K\vec{n}_{i}\cdot\vec{n}_{j}-\sum_{ij}\vec{D}_{ij}\cdot
(\vec{n}_i\times\vec{n}_j). }
Here $\vec n$ denotes the normal vector to the membrane embedded in $3D$-space.
The first term is a bending rigidity term and $K$ is the bending rigidity modulus. 
The second term is a random spontaneous curvature term with a Gaussian
probability distribution for $\vec{D}_{ij}$ with variance
\eqn\Ddis{<\vec{D}_{ij}^{2}>=\Gamma^{2}.}

In particular, it does not satisfy ``physical'' constraints of a spontaneous
curvature which would have been induced by crumpling.
Within a mean field study, it was concluded in [\xref\BMPS,\xref\ACB] 
that the model has a wrinkled 
phase in the $(K,\Gamma)$-plane with nonzero spin glass order parameter $q\neq 0$.
This is to be contrasted with our results where the existence of such a phase 
was crucially requiring the ``physical'' constraint on the disorder variable.
However it is not yet clear whether our conclusions for a quenched
random rigidity are applicable to the quenched random spontaneous curvature case.  
To study the folding of the triangular lattice with random spontaneous curvature, 
we would need to go to a three-dimensional embedding space. 
One possibility is to introduce disorder in the 96-vertex model of ref.\BDIGG.
This is left for future study.

\vskip 1.cm

\leftline{\bf Acknowledgments}

One of the authors (S.M.) thanks Dr.Y.Ozeki for suggesting several important 
references about the cluster variational method.
This work is partially supported by the JSPS Fellowship for Junior Scientists.
We thank Henri Orland for a critical reading of the manuscript.

\listrefs

\end